\documentclass{amsproc}
\usepackage{epsfig}
\theoremstyle{definition}

\theoremstyle{remark}

\numberwithin{equation}{section}

\renewcommand{\a}{\alpha}
\renewcommand{\b}{\beta}

\newcommand{\e}{\epsilon}

\def\a{\alpha}\def\b{\beta}\def\l{\lambda}
\def\s{\sigma}

\newcommand{\p}[1]{(\ref{#1})}    
\newcommand{\be}{\begin{equation}}
\newcommand{\ee}{\end{equation}}
\newcommand{\bea}{\begin{eqnarray}}
\newcommand{\eea}{\end{eqnarray}}
\newcommand{\nn}{\nonumber \\}         
\newcommand{\rep}[1]{\mathbf{#1}}   
\newcommand{\trsp}{{\mathrm T}}

\newcommand{\ii}{\mathrm{i}}

\newcommand{\bbR}{\mathbb{R}}
\newcommand{\bbZ}{\mathbb{Z}}

\newcommand{\bbC}{\mathbb{C}}

\newcommand{\dcy}{d}               % dimension of cycle
\newcommand{\newa}{a}               % dimension of cycle
\newcommand{\newf}{h}               % dimension of cycle

\begin{document}

\title{Branes, Calibrations and Supergravity}

% Remove or comment out any unused author tags.
% author one information
\author{Jerome P. Gauntlett}
\address{Department of Physics, Queen Mary, University of London, Mile
End Rd, London, E1 4NS, U.K. }
%\curraddr{}
\email{J.P.Gauntlett@qmul.ac.uk}
\thanks{
The author thanks Bobby Acharya, Nakwoo Kim, Dario Martelli, 
Stathis Pakis and Daniel Waldram for 
enjoyable collaborations upon which some of these notes are based.}

% author two information
%\author{}
%\address{}
%\curraddr{}
%\email{}
%\thanks{}

% Use this \subjclass if you are using amsproc version 2.0 (December 1999).
\subjclass[2000]{}
% Use this one if you are using an older version of amsproc.
%\subjclass{}
\date{March 2003}

\begin{abstract}
These notes are based on lectures given at the Clay School on
 Geometry and String Theory, Isaac Newton Institute, Cambridge, 25 March - 19 
April 2002. They attempt to provide an elementary and somewhat
self contained discussion of 
the construction of supergravity solutions describing branes wrapping
calibrated cycles, emphasising the geometrical aspects and focusing
on D=11 supergravity. Following a discussion of the role of special holonomy backgrounds in D=11 
supergravity, the basic membrane and fivebrane solutions are reviewed 
and the connection with the AdS/CFT correspondence is made. 
The world-volume description of branes is introduced and used to argue
that branes wrapping calibrated cycles in special holonomy manifolds
preserve supersymmetry.
The corresponding supergravity solutions are constructed first in
an auxiliary gauged supergravity theory which is obtained via Kaluza-Klein reduction.
\end{abstract}

\maketitle

\section{Introduction}

Supergravity theories in D=10 and D=11 spacetime dimensions play an important
role in string/M-theory since they describe the low-energy dynamics.
There are five different string theories all in D=10. At low energies 
the type IIA and type IIB string theories give rise to type IIA 
and type IIB supergravity, respectively, while the type I, and 
the two heterotic string theories all give rise to type I supergravities.
The five string theories are all related to each other, 
possibly after compactification,
by various dualities. There are also dualities which relate string theory
to M-theory, which resides in D=11. 
M-theory is much less understood than string theory, but 
one of the most important things that is known about it, is that its low-energy 
effective action is given by D=11 supergravity.

Solutions to the supergravity equations of motion, particularly those that preserve 
some supersymmetry, are of interest for many reasons. One reason is that they are
useful in studying compactifications from D=10 or 11 down to a lower dimensional spacetime.
By compactifying down to four spacetime dimensions, for example, one might hope
to make contact with particle physics phenomenology. 
In addition to strings, it is known that  string theory has a rich spectrum of other
extended objects or ``branes''. Indeed supergravity solutions can be constructed
describing the geometries around such branes, and these provide a very important
description of the branes. Similarly, there are membrane and 
fivebrane solutions of D=11 supergravity, which will be reviewed later, which implies
that M-theory contains such branes. An important application of brane and more general
intersecting brane solutions is that they can be used to effectively
study the quantum properties of black holes.

Supergravity solutions also 
provide powerful tools to study quantum field theories. The most significant example
is Maldacena's celebrated AdS/CFT correspondence \cite{Maldacena:1998re}
which conjectures that string/M-theory
on certain supergravity geometries that include anti-de-Sitter (AdS) space factors,
is equivalent to certain conformally invariant quantum field theories (CFTs). 
The supergravity approximation to string/M-theory allows one to calculate
highly non-trivial information about the conformal field theories. 

The AdS/CFT correspondence is truly remarkable. On the one hand it states that
certain quantum field theories, that {\it a priori} have nothing to do with gravity,
are actually described by theories of quantum gravity (string/M-theory).
Similarly, and equally surprising, it also states that quantum gravity on certain 
geometries, is actually quantum field theory. As a consequence
much effort has been devoted to further understanding and generalising the correspondence. 

The basic AdS/CFT examples arise from studying 
the supergravity solutions describing planar branes in flat space, in the
``near horizon limit''. Roughly speaking, this is the limit close
to the location of the brane. Here we shall discuss more general
supergravity solutions that describe branes that are partially wrapped on various
calibrated cycles within special holonomy manifolds. We will construct explicit solutions
in the near horizon limit, which is sufficient for applications to the  AdS/CFT correspondence.

To keep the presentation simple, we will mostly restrict our discussion to D=11 supergravity.
In an attempt to make the lectures accessible to both maths
and physics students we will emphasise the geometrical aspects and de-emphasise the
quantum field theory aspects. To make the discussion somewhat self contained, we 
begin with some basic material; it is hoped that the discussion is not
too pedestrian for the physics student and not too vague for the maths student!

We start with an introduction to D=11 supergravity, defining the notion of a bosonic 
solution of D=11 supergravity that preserves supersymmetry. We then describe why manifolds
with covariantly constant spinors, and hence with special holonomy, are important. Following
this we present the geometries describing planar membranes and fivebranes. 
These geometries have horizons, and in the near horizon limit we obtain geometries 
that are products of $AdS$ spaces with spheres, which leads to a 
discussion of the basic AdS/CFT examples.
 
To motivate the search for new AdS/CFT examples we first
introduce the world-volume description of branes.
Essentially, this is an approximation that treats the branes as ``probes'' propagating
in a fixed background geometry. We define calibrations and calibrated cycles,
and explain why such probe-branes wrapping calibrated cycles in special holonomy 
manifolds preserve supersymmetry. 
The aim is then to construct supergravity solutions describing such
wrapped branes after including the back-reaction of the branes on the special
holonomy geometry. 

The construction of these supergravity solutions is a little subtle. 
In particular, the solutions are first constructed 
in an auxiliary gauged supergravity theory. We will focus, for illustration, 
on the geometries corresponding to wrapped fivebranes. 
For this case the solutions are first found in $SO(5)$ gauged supergravity in D=7.
This theory arises from the consistent truncation of the dimensional reduction of 
D=11 supergravity on a four-sphere, as we shall discuss. This means that any
solution of the D=7 supergravity theory automatically gives a solution 
of D=11 supergravity. 
We will present several details of the construction of the solutions describing
fivebranes wrapping SLAG 3-cycles, and summarise more briefly the other
cases. We also comment on some aspects of the construction of the solutions for 
wrapped membranes and D3-branes of type IIB supergravity.

We conclude with a discussion section that outlines some open problems
as well as a brief discussion of other related work on the construction
of wrapped NS-fivebranes  of type IIB supergravity.

\section{D=11 supergravity}

The bosonic field content of D=11 supergravity 
\cite{Cremmer:1978km} consists of a metric, 
$g$, and a three-form $C$ with four-form field strength $G=dC$ 
which live on a D=11 manifold which we take to be spin. The signature is 
taken to be mostly plus, $(-,+,\dots,+)$. In addition the
theory has a fermionic gravitino, $\psi_\mu$.
The action with $\psi_\mu=0$ is given by
\begin{equation}
S=\frac{1}{2\kappa^2}\int d^{11} x {\sqrt{-g}}R
-\frac{1}{2}G\wedge *G - \frac{1}{6}C\wedge G\wedge G~,
\end{equation}
and thus the bosonic equations of motion, including the Bianchi identity for the four-form, 
are
\begin{eqnarray}
R_{\mu\nu}&=&\frac{1}{12}(G^2_{\mu\nu}
-\frac{1}{12}g_{\mu\nu}G^2)\nn 
d*G+\frac{1}{2}G\wedge G&=&0\nn
dG&=&0~,
\end{eqnarray}
where $G^2_{\mu\nu}= G_{\mu\s_1\s_2\s_3}G{_{\nu}}{^{\s_1\s_2\s_3}}$ and
$G^2=G_{\s_1\s_2\s_3\s_4}G^{\s_1\s_2\s_3\s_4}$, with  $\mu,\nu,\s =0,1,\dots, 10$. 
The theory is invariant under supersymmetry transformations
whose infinitesimal form is given schematically by
\begin{eqnarray}
\label{schemvar}
\delta g&\sim& \epsilon\psi\nn
\delta C&\sim& \epsilon\psi\nn
\delta\psi&\sim&\hat\nabla\epsilon +\epsilon\psi\psi~,
\end{eqnarray}
where the spinor $\epsilon$ parametrises the variation and the
connection $\hat\nabla$ will be given shortly.

Of primary interest are bosonic solutions to the equations of
motion that preserve at least one supersymmetry. These are solutions to 
the equations of motion with $\psi=0$ 
which are left inert under a supersymmetry 
variation. From \p{schemvar} we see that $\delta g=\delta C=0$ trivially, 
and hence we seek solutions to the equations of motion
that admit non-trivial solutions to the equation $\hat \nabla \epsilon=0$.

As somewhat of an aside we mention a potentially confusing point. {}For the 
theory to be supersymmetric, it is necessary that all of the fermions are
Grasssmann odd (anti-commuting) spinors. 
However, since the only place that fermions enter into bosonic supersymmetric
solutions is via $\hat \nabla \epsilon=0$ and since this is
linear in $\e$ we can, and will, take $\e$ to be a commuting
(i.e. ordinary) spinor from now on. The Grassmann odd character of the
fermions is certainly important in the quantum theory, but this will not concern us
here.

To be more precise about the connection $\hat\nabla$
let us introduce some further notation.
We will use the convention that $\mu,\nu,\dots$ are co-ordinate indices and
$\alpha,\beta,\dots$ are tangent space indices, i.e., indices with respect to an orthonormal frame.
The D=11 Clifford algebra, $Cliff(10,1)$, is generated by 
gamma-matrices $\Gamma^\a$ satisfying the algebra 
\begin{equation}
\Gamma^\a \Gamma^\b+\Gamma^\a \Gamma^\b=2 \eta^{\a\b}~,
\end{equation}
with $\eta=diag(-1,1,\dots,1)$.
We will work in a representation where the gamma-matrices are real
$32\times 32$ matrices acting on real 32 component spinors, with
$\Gamma_0\Gamma_1\dots \Gamma_{10}=+1$. Recall that
$Spin(10,1)$  is generated by 
\begin{equation}
\frac{1}{4}\Gamma^{\a\b}\equiv\frac{1}{8}\left(\Gamma^\a\Gamma^\b-
\Gamma^\a\Gamma^\b\right)~,
\end{equation}
and here we have introduced the notation that $\Gamma^{\a_1\dots\a_p}$ is an
anti-symmetrised product of $p$ gamma-matrices. The charge conjugation matrix
is defined to be $\Gamma_0$ and $\bar\e\equiv \e^\trsp\Gamma_0$.

We can now write the condition for a bosonic configuration to
preserve supersymmetry as
\begin{equation}\label{killspin}
\hat \nabla_\mu \epsilon\equiv
\nabla_\mu\epsilon+\frac{1}{288}[\Gamma{_\mu}{^{\nu_1\nu_2\nu_3\nu_4}}
-8\delta{_\mu^{\nu_1}}\Gamma^{\nu_2\nu_3\nu_4}]G_{\nu_1\nu_2\nu_3\nu_4}\e=0~,
\end{equation}
where $\nabla_\mu\epsilon$ is the usual covariant derivative on the spin bundle
\begin{equation}
\nabla_\mu\epsilon=(\partial_\mu+\frac{1}{4}\omega_{\mu\a\b}\Gamma^{\a\b})\epsilon~.
\end{equation}
Observe that the terms involving the four-form in  
\p{killspin} imply that $\hat\nabla$ takes values in 
the Clifford algebra and not just the Spin subalgebra. 
This is the typical situation in
supergravity theories but there are exceptions, such as
type I supergravity, where the connection takes values
in the spin subalgebra and has totally anti-symmetric torsion \cite{Strominger:1986uh}.

Non-trivial solutions to \p{killspin} are called Killing spinors. 
The nomenclature is appropriate since if $\e^i$, $\e^j$ are Killing spinors
then $K^{ij\mu}\equiv\bar\e^i\Gamma^\mu\epsilon^j$ are Killing vectors. 
To see this, first 
define $\Omega^{ij}_{\mu\nu}
=\bar\e^i \Gamma_{\mu\nu}\e^j$ and $\Sigma^{ij}_{\mu_1\dots \mu_5}=
\bar\e^i\Gamma_{\mu_1\dots \mu_5}\e^j$. Then use \p{killspin}
to show that \cite{Gauntlett:2002fz}
\begin{equation}
\nabla_\mu K^{ij}_\nu=\frac{1}{6}\Omega^{ij}{^{\s_1\s_2}}G_{\s_1\s_2\mu\nu}
+\frac{1}{6!}\Sigma^{ij}{^{\s_1\s_2\s_3\s_4\s_5}}*G_{\s_1\s_2\s_3\s_4\s_5\mu\nu}~,
\end{equation}
and hence in particular $\nabla_{(\mu} K^{ij}_{\nu)}=0$. 
It can also be shown that the ``diagonal'' Killing vectors $K^{ii}$,
for each Killing spinor $\e^i$, are either time-like or 
null \cite{MR2002h:53082}.
The zeroth component
of these vectors in an orthonormal frame are given by $(\e^i)^\trsp\e^i$, and are 
clearly non-vanishing if and only if $\e^i$ is, and 
hence so is $K$ itself. 

It is useful to know under what conditions a
geometry admitting a Killing spinor will also solve the equations of
motion. In the case when there is a time-like Killing spinor,
i.e. a Killing spinor whose corresponding Killing vector is time-like,
it was proved in \cite{Gauntlett:2002fz} that the geometry will
solve all of the equations of motion providing that $G$ satisfies the Bianchi
identity $dG=0$ and the four-form equation of motion 
$d*G+1/2G\wedge G=0$. If all of the
Killing spinors are null, it is necessary, in addition, to demand
that just one component of the Einstein equations are satisfied 
\cite{Gauntlett:2002fz}.

Note that given the value of a Killing spinor at a point, the connection
defines a Killing spinor everywhere, via parallel transport. Also, as the
Killing spinor equation is linear, the Killing spinors form a vector
space whose dimension $n$ can, in principle, be from $1,\dots, 32$. 
The fraction of preserved supersymmetry is then $n/32$. Although solutions
are known preserving many fractions of supersymmetry, it is not yet known if
all fractions can occur (for some recent
speculations on this issue 
see \cite{Duff:2003ec}). A general characterisation of the most
general supersymmetric geometries preserving one time-like Killing spinor
is presented in \cite{Gauntlett:2002fz}. It was shown that the geometry
is mostly determined by a ten dimensional manifold orthogonal to
the orbits of the Killing vector that admits an $SU(5)$-structure
with rather weakly constrained intrinsic torsion. 
The analogous analysis for null Killing spinors has not yet been 
carried out. A complete
classification of maximally supersymmetric solutions preserving all 32 supersymmetries is presented in \cite{Figueroa-O'Farrill:2002ft}.

Most of our considerations will be in the context of D=11 supergravity, but
it is worth commenting on some features of M-theory that embellish
D=11 supergravity. Firstly, the flux $G$, which is unconstrained in D=11
supergravity, satisfies a quantisation condition in M-theory.
Introducing the Planck length $l_p$, via
\be
2\kappa^2\equiv (2\pi)^8 l_p^9~,
\ee
for M-theory on a D=11 spin manifold $Y$ we have \cite{Witten:1997md}
\be\label{gquantcon}
\frac{1}{(2\pi l_p)^3}G-\frac{\lambda}{2}\in H^4(Y,\bbZ)~,
\ee
where $\lambda(Y)=p_1(Y)/2$ with $p_1(Y)$ the first Pontryagin class of $Y$, given below.
Note that since $Y$ is a spin manifold, $p_1(Y)$ is divisible by two. Actually, more generally
it is possible to consider M-theory on unorientable manifolds admitting pinors and some
discussion can be found in \cite{Witten:1997md}.

A second point is that the low-energy effective
action of M-theory is given by D=11 supergravity supplemented by an
infinite number of higher order corrections. It is not yet known
how to determine almost all of these corrections, but there is one 
important exception. Based on
anomaly considerations it has been shown that the equation of motion for the
four-form $G$ is modified, at next order, by \cite{Vafa:1995fj,Duff:1995wd}
\be
d*G+\frac{1}{2}G\wedge G=-\frac{(2\pi l_p)^6}{192}
\left(p_1^2-4 p_2\right)~,
\ee
where the first and second Pontryagin forms are given by
\be
p_1=-\frac{1}{8\pi^2}tr R^2~,\qquad p_2=-\frac{1}{64\pi^4}tr R^4
+\frac{1}{128\pi^4}(tr R^2)^2~.
\ee
At the same order there are other corrections
to the equations of motion and also to the supersymmetry variations, but these have
not yet been determined. Thus it is not yet known how to fully incorporate this
correction consistently with supersymmetry but nevertheless it does
have important consequences (see e.g. \cite{Sethi:1996es}). 
As this correction will not play a role in subsequent discussion, we will ignore it.

In the next sub-sections we will review two basic classes of supersymmetric solutions
to D=11 supergravity. The first class are special holonomy manifolds and the
second class are the membrane and fivebrane solutions.

\subsection{Special Holonomy}
{}First consider supersymmetric solutions that have vanishing four-form flux $G$, where
things simplify considerably. The equations of motion and the Killing spinor
equation then become
\bea
R_{\mu\nu}&=&0\nn
\nabla_\mu\e&=&0~.
\eea
That is, Ricci-flat manifolds with covariantly constant spinors. The second
condition implies that the manifolds have special holonomy.
To see this, observe that it implies the integrability condition
\be\label{intcond}
[\nabla_\mu,\nabla_\nu]\e=\frac{1}{4}R_{\mu\nu\a\b}\Gamma^{\a\b}\e=0~.
\ee
The subgroup of $Spin(10,1)$ generated by $R_{\mu\nu\a\b}\Gamma^{\a\b}$
gives the restricted holonomy group $H$. Thus \p{intcond} implies that
a Killing spinor must be invariant under $H$. i.e.
it must be a singlet under the decomposition
of the $\rep{32}$ spinor representation of $Spin(10,1)$ into 
$H$ representations, and this constrains the possible holonomy
groups $H$ that can arise.

Of most interest to us here are geometries $\bbR^{1,10-d}\times M_d$,
which are the direct product of $(11-d)$-dimensional 
Minkowski space, $\bbR^{1,10-d}$, with 
a $d$-dimensional Riemannian Manifold $M_d$, which we mostly
take to be simply connected.
({}For a discussion of supersymmetric solutions with
Lorentzian special holonomy, see \cite{MR2002h:53082,Figueroa-O'Farrill:1999tx}). 
The possible holonomy groups of the Levi-Civita connection on manifolds $M_d$ 
admitting covariantly constant spinors is well known, and we now 
briefly summarise the different cases.

{\bf $Spin(7)$-Holonomy}: In $d=8$ there are Riemannian
manifolds with $Spin(7)$ holonomy. 
These have a no-where vanishing self-dual Cayley four-form $\Psi$ whose
components in an orthonormal frame can be taken as 
\begin{equation}
\label{Psidef}
\begin{split}
   \Psi &= e^{1234}+e^{1256}+e^{1278}+e^{3456}+e^{3478}+e^{5678} \\
      &\qquad + e^{1357}-e^{1368}-e^{1458}-e^{1467}-e^{2358}
          -e^{2367}-e^{2457}+e^{2468}~,
\end{split}
\end{equation}
where e.g. $e^{1234}=e^1\wedge e^2\wedge e^3\wedge e^4$.
The Cayley four-form is covariantly constant for $Spin(7)$ manifolds
and this is equivalent to $\Psi$ being closed:
\be
d\Psi =0~.
\ee
$Spin(7)$ holonomy manifolds have a single covariantly constant
chiral $Spin(8)$ spinor, which we denote by $\rho$. Moreover, the Cayley
four-form can be constructed as a bi-linear in $\rho$:
\be
\Psi_{mnpq}=-\bar\rho\gamma_{mnpq}\rho~,
\ee
where here $m,n,p,q=1,\dots ,8$. {}For more discussion on the
spinor conventions for this case and those below, see appendix B of
\cite{Gauntlett:2003cy}.

{\bf $G_2$-Holonomy}: In $d=7$ there are Riemannian
manifolds with $G_2$ holonomy. These have
a no-where vanishing associative three-form $\phi$ whose components in
an orthonormal frame can be taken as 
\begin{equation}
\label{g2form}
   \phi = e^{246}-e^{235} - e^{145} - e^{136}
      + e^{127} + e^{347} + e^{567}~.
\end{equation}
The three-form is covariantly constant and this is in fact equivalent
to the conditions
\begin{equation}
d\phi=d*\phi=0~.
\end{equation}
These geometries possess a single covariantly constant minimal $d=7$ spinor $\rho$.
The associative three-form can be constructed from $\rho$ via
\begin{equation}
   \phi_{mnp} = - \ii \bar{\rho} \gamma_{mnp} \rho~.
\end{equation}

{\bf $SU(n)$-Holonomy}: In $d=2n$ there are Calabi-Yau $n$-folds ($CY_n$) with
$SU(n)$ holonomy. The cases relevant for D=11 supergravity have
$n=2,3,4,5$. Calabi-Yau manifolds are complex manifolds, 
with complex structure $J$, and admit a no-where vanishing holomorphic 
$(n,0)$-form $\Omega$. The K\"ahler form, which we also denote by $J$,
is obtained by lowering an index on the complex structure. In an orthonormal
frame we can take
\bea
J&=&e^{12}+e^{34}+\dots+e^{(2n-1)(2n)}\nn
\Omega&=&(e^1+ie^2)(e^3+ie^4)\dots(e^{2n-1}+ie^{2n})~.
\eea
Both $J$ and $\Omega$ are covariantly constant and this is equivalent
to the vanishing of the exterior derivative of the K\"ahler-form and 
the holomorphic $(n,0)$-form:
\be
dJ=d\Omega=0~.
\ee
These manifolds have a covariantly constant complex chiral spinor $\rho$.
The complex conjugate of this spinor is also covariantly constant.
For $n=2,4$ the conjugate spinor has the same chirality, while for
$n=3,5$ it has the opposite chirality. 
$J$ and $\Omega$ can be written in terms of the spinor $\rho$
as
\bea
J_{mn}&=&\ii\rho^\dag\gamma_{mn}\rho\nn
\Omega_{m_1\dots m_{2n}}&=&\rho^\trsp\gamma_{m_1\dots m_{2n}}\rho~.
\eea

{\bf $Sp(n)$-Holonomy}: 
In $d=4n$ there are hyper-K\"ahler $n$-manifolds ($HK_n$) with $Sp(n)$
holonomy. The cases relevant for D=11 supergravity have $n=1,2$.
These admit three covariantly constant complex structures $J^{a}$ satisfying
the algebra of the imaginary quaternions
\begin{equation}
J^{a}\cdot J^{b}=-\delta^{ab} +\e^{abc} J^{c}~.
\end{equation}
If we lower an index on the $J^{a}$ we obtain
three K\"ahler-forms and the condition for  $Sp(n)$-holonomy is equivalent
to their closure:
\be
dJ^{a}=0~.
\ee
Note that when $n=1$, since $Sp(1)\cong SU(2)$, four dimensional
hyper-K\"ahler manifolds are equivalent to Calabi-Yau two-folds.
{}From the $CY_2$ side,
the extra two complex structures are obtained from the holomorphic 
two-form via $\Omega=J^2+iJ^1$. The remaining case of interest for D=11
supergravity is eight-dimensional hyper-K\"ahler manifolds when $n=2$.
In this case, in an orthonormal frame we can take the three K\"ahler forms to be given
by
\begin{equation}
\begin{aligned}\label{jays}
   J^1 &= e^{12} + e^{34} + e^{56} + e^{78}  \\
   J^2 &= e^{14} + e^{23} + e^{58} + e^{67}  \\
   J^3 &= e^{13} + e^{42} + e^{57} + e^{86}~.
\end{aligned}
\end{equation}
Each complex structure $J^a$ has a corresponding holomorphic
$(4,0)$ form given by
\begin{equation}
\begin{aligned}\label{omegas}
   \Omega^1 &= \tfrac{1}{2}J^3\wedge J^3 - \tfrac{1}{2}J^2\wedge J^2
      + \ii J^2\wedge J^3  \\
   \Omega^2 &= \tfrac{1}{2}J^1\wedge J^1 - \tfrac{1}{2}J^3\wedge J^3
      + \ii J^3\wedge J^1  \\
   \Omega^3 &= \tfrac{1}{2}J^2\wedge J^2 - \tfrac{1}{2}J^1\wedge J^1
      + \ii J^1\wedge J^2~.
\end{aligned}
\end{equation}
These manifolds have three covariantly constant $Spin(8)$ spinors of the 
same chirality $\rho_a$, $a=1,2,3$.
The three K\"ahler forms can be constructed as
\begin{equation}
\begin{aligned}
   J^1_{mn} &= - \bar\rho_{2}\gamma_{mn}\rho_{3} \\
   J^2_{mn} &= - \bar\rho_{3}\gamma_{mn}\rho_{1} \\
   J^3_{mn} &= - \bar\rho_{1}\gamma_{mn}\rho_{2}~.
\end{aligned}
\end{equation}

In addition to these basic irreducible examples we can also consider
$M_d$ to be the direct product of two manifolds. A rather trivial
possibility is to consider the product of one of the above manifolds
with a number of flat directions. Two non-trivial
possibilities are to consider the product $CY_3\times CY_2$ with
$SU(3)\times SU(2)$ holonomy, or the product $CY_2\times CY_2'$ with
$SU(2)\times SU(2)$ holonomy.

We have summarised the possibilities in table 1. 
We have also recorded the amount
of D=11 supersymmetry preserved by geometries of the form 
$\bbR^{1,10-d}\times M_d$. As
we noted this corresponds to the total number of singlets in the decomposition
of $\rep{32}$ of $Spin(10,1)$ into representations of $H$. Let us illustrate the
counting for the $d=8$ cases. The spinor representation $\rep{32}$ of 
$Spin(10,1)$ decomposes into $Spin(2,1)\times Spin(8)$ representations as
\begin{equation}\label{decomp}
\rep{32}\to (\rep{2},\rep{8_+})+(\rep{2},\rep{8}_-)~,
\end{equation}
where the subscripts refer to the chirality of the two spinor
representations of $Spin(8)$. When $M_8$ is a  $Spin(7)$-manifold
we have the further decomposition under $Spin(7)\subset Spin(8)$
\begin{equation}
\rep{8}_+\to \rep{7}+\rep{1},\qquad \rep{8}_-\to \rep{8}~.
\end{equation}
The singlet corresponds to the single covariantly constant, $Spin(7)$
invariant, spinor on the $Spin(7)$-manifold discussed above. From
\p{decomp} we see that this gives rise to two preserved supersymmetries
and that they transform as a minimal two real component spinor of $Spin(2,1)$.
This is also described as preserving $N=1$ supersymmetry in 
D=3 spacetime dimensions corresponding to the $\bbR^{1,2}$ factor.
When $M_8$ is Calabi-Yau under $SU(4)\subset Spin(8)$ we
have 
\begin{equation}
\rep{8}_+\to \rep{6}+\rep{1}+\rep{1},\qquad \rep{8}_-\to \rep{4}+\rep{\bar4}~.
\end{equation}
The two singlets combine to form the complex covariantly constant spinor
on $CY_4$ mentioned above. In this case four supersymmetries are preserved, 
transforming as two minimal
spinors of $Spin(2,1)$, or $N=2$ supersymmetry in D=3. When $M_8$ is 
hyper-K\"ahler, under $Sp(2)\subset Spin(8)$ we have
\begin{equation}
\rep{8}_+\to \rep{5}+\rep{1}+\rep{1}+\rep{1},\qquad \rep{8}_-\to \rep{4}+\rep{4}~,
\end{equation}
and six supersymmetries are preserved, or $N=3$ in D=3.
Similarly, when $M_8$ is the product of two Calabi-Yau two-folds 
eight supersymmetries are preserved, or $N=4$ in D=3. If we also allow tori, then 
when $M_8$ is the product of a Calabi-Yau two-fold with $T^4$, sixteen
supersymmetries are preserved, or $N=8$ in D=3, while the simple case of 
$T^8$ preserves all thirty two supersymmetries or $N=16$ in D=3.

\begin{table}[!th]
\begin{center}
\setlength{\tabcolsep}{0.45em}
\begin{tabular}{|c|c|c|}
\hline
dim($M$) & Holonomy&Supersymmetry\\
\hline\hline
10 & $SU(5)$     & 2 \\
10 & $SU(3)\times SU(2)$     & 4\\
8 & $Spin(7)$     & 2 \\
8 & $SU(4)$     & 4 \\
8 & $Sp(2)$     & 6 \\
8 & $SU(2)\times SU(2)$     & 8 \\
7 & $G_2$     & 4 \\
6 & $SU(3)$     & 8 \\
4 & $SU(2)$     & 16 \\
\hline
\end{tabular}
\end{center}
\caption{Manifolds of special holonomy and the corresponding 
amount of preserved supersymmetry.}
\label{green}
\end{table}

An important way to make contact with four-dimensional physics is to 
consider geometries of the form $\bbR^{1,3}\times M_7$ with $M_7$ compact.
If we choose $M_7$ to be $T^7$ then it preserves all
32 supersymmetries or $N=8$  supersymmetry in D=4 spacetime dimensions. 
$T^3\times CY_2$ preserves 16 
supersymmetries or $N=4$ in D=4, $S^1\times CY_3$ 
preserves 8 supersymmetries or $N=2$ in D=4
and $G_2$ preserves four supersymmetries or $N=1$ in D=4.

$N=1$ supersymmetry in four spacetime dimensions
has many attractive
phenomenological features and this is the key reason for the recent interest
in manifolds with $G_2$ holonomy. As discussed in Acharya's lectures at this
school,
it is important to emphasise that the most interesting examples
from the physics point of view are not complete. In addition one
can use non-compact $G_2$ holonomy manifolds very effectively to study various
quantum field theories in four spacetime dimensions (see e.g. \cite{Atiyah:2001qf}).

Another important class of examples is to consider $d=7$ manifolds of the
form $S^1/Z_2 \times CY_3$ where the $Z_2$ action has two fixed planes. 
It can be shown that the orbifold breaks a further one
half of the supersymmetries and one is again left with four supersymmetries
in four spacetime dimensions. These configurations are related to the
the strongly coupled limit of heterotic string theory compactified on $CY_3$
\cite{Horava:1996qa,Horava:1996ma}.

In summary, when $G=0$, geometries of
the form $\bbR^{1,10-d}\times M_d$ preserve supersymmetry when $M_d$ 
admits covariantly constant spinors and hence has special holonomy. 
For physical applications, $M_d$ need not be compact nor complete. 
In the next section we will consider the basic solutions with $G\ne 0$,
the fivebrane and the membrane solutions.

\subsection{Membranes and Fivebranes}
The simplest, and arguably the most important supersymmetric
solutions with non-vanishing four-form are the
membrane and fivebrane solutions. {}Further discussion can be found
in e.g. \cite{Stelle:1998xg}.

The fivebrane geometry is given by
\bea\label{fivegeom}
ds^2&=&H^{-1/3}\left[d\xi^i d\xi^j \eta_{ij}\right]
+H^{2/3}\left[dx^I dx^I\right]\nn
G_{I_1I_2I_3I_4}&=&-c\epsilon_{I_1I_2I_3I_4J}\partial_JH~, \qquad c=\pm 1~,
\eea
where $i,j=0,1,\dots 5$, $I=1,\dots,5$ and $H=H(x^I)$. 
This geometry preserves 1/2 of the supersymmetry. In the obvious orthonormal
frame $\{H^{-1/6}d\xi^i,H^{1/3}dx^I\}$, the 16 Killing spinors are given by 
\be
\epsilon=H^{-1/12}\e_0~,
\ee
where $\e_0$ is a constant spinor, and satisfy
\be
\Gamma^{012345}\e=c\e~.
\ee
Since $\Gamma^{012345}$ squares to unity and is traceless, we conclude
that the geometry admits 16 independent Killing spinors.

This geometry satisfies the equations of motion providing that
we impose the Bianchi identity for $G$ which implies that
$H$ is harmonic. 
%It is convenient at this point to
%define the D=11 planck length, $l_p$, via
%$2\kappa^2\equiv (2\pi)^8 l_p^9$. 
If we take $H$ to have a single centre
\be\label{m5singcent}
H=1+\frac{\a_5N}{r^3},\qquad r^2=x^Ix^I~,
\ee
with $N$ positive and $\a_5 =\pi l_p^3$, then the solution
carries $cN$ units of quantised magnetic four-form flux 
\be
\frac{1}{(2\pi l_p)^3}\int_{S^4} G=cN~,
\ee
with $N$ a positive integer, consistent with \p{gquantcon}.
When $c=+1$ the solution describes $N$ coincident fivebranes, 
that are oriented along the  $d\xi^0\wedge d\xi^1\wedge\dots d\xi^5$ 
plane. When $c=-1$ the solution describes $N$ coincident
anti-fivebranes. Roughly speaking, the fivebranes can be thought of
as being located at $r=0$, where the solution appears singular.
However, this is in fact a regular horizon and moreover, 
it is possible to analytically continue to obtain a completely 
non-singular geometry \cite{Gibbons:1995vm}. Thus 
it is not possible to say exactly where the fivebranes are located.

In the directions transverse to the fivebrane the metric 
becomes asymptotically flat. We can thus calculate the ADM mass
per unit volume, or tension, and we find
\be
Tension =NT_5~,\qquad T_5=\frac{1}{(2\pi)^5 l_p^6}~,
\ee
where $T_5$ is the tension of a single fivebrane (for a careful
discussion of numerical co-efficients appearing in $T_5$ and
the membrane tension $T_2$ below, see \cite{deAlwis:1996ez}).
It is possible to show that the supersymmetry algebra actually
implies that the tension of the fivebranes is fixed by
the magnetic charge. This ``BPS'' condition is equivalent to
the geometry preserving 1/2 of the supersymmetry.
Note also that if $H$ is taken to be a multi-centred harmonic function then we
obtain a solution with the $N$ co-incident fivebranes separated.

It is interesting to examine the near horizon limit of the geometry 
of $N$ coincident fivebranes, when $r\approx 0$. By dropping 
the one from the harmonic function in \p{m5singcent} we get
\bea
ds^2&=&\frac{r}{(\a_5 N)^{1/3}}\left[d\xi^i d\xi^j \eta_{ij}\right]
+\frac{(\a_5 N)^{2/3}}{r^2}\left[dr^2 +r^2d\Omega_4\right]~.
\eea
where $d\Omega_4$ is the metric on the round four-sphere.
After a co-ordinate transformation we can rewrite this as
\bea\label{fivenear}
ds^2&=&R^2\left[
\frac{d\xi^i d\xi^j \eta_{ij}+d\rho^2}{\rho^2}\right]
+\frac{R^2}{4}d\Omega_4~,
\eea
which is just $AdS_7\times S^4$, in Poincar\'e co-ordinates,
with the radius of the $AdS_7$ given by
\be
R=2(\pi N)^{1/3}l_p~.
\ee
There are still $N$ units of flux on the four-sphere.
This geometry is in fact a solution to the equations of motion that
preserves all 32 supersymmetries. A closely related fact is that
the Lorentz symmetry $SO(5,1)$ of the fivebrane solution has been
enhanced to the conformal group $SO(6,2)$. The interpretation of this
fact and the $SO(5)$ isometries of the four-sphere will be discussed in the next
section.
Before doing so, we introduce the membrane solution.

The membrane geometry is given by
\bea\label{memgeom}
ds^2&=&H^{-2/3}\left[d\xi^i d\xi^j \eta_{ij}\right]
+H^{1/3}\left[dx^I dx^I\right]\nn
C&=&cH^{-1}d\xi^0\wedge d\xi^1 \wedge d\xi^3,\qquad c=\pm1~,
\eea
with, here, $i,j=0,1,2$, $I=1,\dots 8$ and $H=H(x^I)$. 
This geometry preserves one half of the supersymmetry. 
Using the orthonormal frame
$\{H^{-1/3}d\xi^i,H^{1/6}dx^I\}$
we find that the Killing spinors are given by
\be
\epsilon=H^{-1/6}\e_0~,
\ee
where $\e_0$ is a constant spinor, and satisfy the constraint
\be\label{m2susycond}
\Gamma^{012}\e=c\e~.
\ee
Since $\Gamma^{012}$ squares to unity and is traceless, we conclude
that the geometry admits 16 independent Killing spinors.

This geometry solves all of the equations of motion providing 
that we impose the four-form equation of motion. This implies
that the function $H$ is harmonic. Now take
$H$ to be
\be\label{m2singcent}
H=1+\frac{\alpha_2N}{r^6},\qquad r^2=x^Ix^I~,
\ee
with $N$ a positive integer and $\alpha_2=32\pi^2l_p^6$.
The solution 
carries $cN$-units of quantised electric four-form charge:
\be
\frac{1}{(2\pi l_p)^6}\int_{S^7} *G=cN~.
\ee
When $c=\pm 1$, the solution describes $N$ coincident (anti-)membranes 
oriented along the $d\xi^0\wedge d\xi^1 \wedge d\xi^3$ plane.
Transverse to the membrane the solution tends to flat space, and we can thus
determine the ADM tension of the membranes. We again find that it is
related to the charge as dictated by supersymmetry
\be
Tension=NT_2~,\qquad T_2=\frac{1}{(2\pi)^2l_p^3}~,
\ee
where $T_2$ is the tension of a single membrane.
If the harmonic function is replaced with a multi-centre 
harmonic function we obtain a solution with the membranes separated.

The solution describing $N$ co-incident membanes appears singular at 
$r\approx 0$, but one can in
fact show that this is a horizon. The solution can be extended across the horizon and
one finds a timelike singularity inside the horizon (see e.g. \cite{Stelle:1998xg}), 
which can be mapped onto a membrane source with tension $T_2$.
To obtain the near horizon geometry, $r\approx 0$, we drop the one in the 
harmonic function \p{m2singcent}, to find, after a co-ordinate
transformation,
\bea
ds^2=R^2\left[
\frac{d\xi^i d\xi^j \eta_{ij}+d\rho^2}{\rho^2}\right]
+4R^2d\Omega_7~,
\eea
which is simply the direct product 
$AdS_4\times S^7$ with the radius of $AdS_4$ given by
\be
R=\left(\frac{N\pi^2}{2}\right)^{1/6}l_p~.
\ee
There are still $N$-units of flux on the seven-sphere. This configuration 
is itself a supersymmetric solution preserving 
all 32 supersymmetries. The $SO(2,1)$ Lorentz symmetry of the membrane 
solution has been enhanced to the conformal group $SO(3,2)$ and the seven
sphere admits an $SO(8)$ group of isometries.

This concludes our brief review of the basic planar membrane and fivebrane solutions.
There is a whole range of more general solutions describing the intersection of
planar membranes and fivebranes, and we refer to the reviews 
\cite{Gauntlett:1997cv,Smith:2002wn} for further details.

\section{AdS/CFT Correspondence}

In the last section we saw that D=11 supergravity admits supersymmetric
solutions corresponding to $N$ co-incident membranes or co-incident
fivebranes, and that in the near horizon limit the metrics become
$AdS_4\times S^7$ or $AdS_7\times S^4$, respectively. 
The famous conjecture of Maldacena \cite{Maldacena:1998re}
states that M-theory on  these backgrounds is 
equivalent to certain conformal field theories in three or 
six spacetime dimensions, respectively. {}For a comprehensive review of
this topic, we refer to \cite{Aharony:1999ti}, but we would like to make a 
few 
comments in order to motivate the construction of the supersymmetric 
solutions of D=11 supergravity presented in later sections. 

The best understood example of the AdS/CFT correspondence 
actually arises in type IIB string theory, so we first pause to 
introduce it. The low-energy limit of type IIB string theory is 
the chiral type IIB supergravity \cite{Schwarz:1983qr,Howe:1984sr}.
The bosonic field content of the supergravity theory consists 
of a metric, a complex scalar, a complex three-form field strength 
and a self-dual five-form field strength. The theory admits 
a 1/2 supersymmetric three-brane, called a D3-brane. The metric of
the corresponding supergravity solution is given by
\be\label{D3geom}
ds^2=H^{-1}\left[d\xi^i d\xi^j\eta_{ij}\right]+H\left[dx^I dx^I\right]~,
\ee
where $i,j=0,1,2,3$, $I,J=1,\dots 6$ and $H=H(x^I)$ is a harmonic function.
If we choose
\be
H=1+\frac{\a_3 N}{r^2}~,
\ee
with $N$ a positive integer, and $\a_3$ some constant with dimensions of length squared,
then the solution corresponds to $N$ co-incident D3-branes. The only
other non-vanishing field is the self-dual five-form and the solution,
for suitably chosen $\a_3$, carries $N$ units of flux when integrated around a five-sphere
surrounding the D3-branes. In the near horizon limit, $r\approx 0$,  
we get $AdS_5\times S^5$, with equal radii. 

The boundary of $AdS_5$ is the conformal compactification
of four-dimensional Minkowski space, $M_4$. 
The AdS/CFT conjecture states that type IIB string theory 
on $AdS_5\times S^5$ is equivalent (dual) 
to ${\mathcal N}=4$ supersymmetric Yang-Mills theory 
with gauge group $SU(N)$ on $M_4$. This quantum field theory is very special 
as it has the maximal amount of supersymmetry that a quantum field theory can have>
Moreover, it is a conformal field theory (CFT), i.e. invariant under the
conformal group.
The AdS/CFT correspondence relates parameters in the field theory with those 
of the string theory on $AdS_5\times S^5$. It turns out that perturbative Yang-Mills theory
can be a good description only when the radius $R$ of the $AdS_5$ is small,
while supergravity is a good approximation only when $R$ is large and
$N$ is large. The fact that these different regimes don't overlap
is a key reason why such seemingly different theories could be equivalent
at all. 

The natural objects to consider in a conformal field theory are correlation
functions of operators. A precise dictionary between operators in the conformal field 
theory and fields (string modes) propagating in $AdS_5$ 
is given in \cite{Gubser:1998bc,Witten:1998qj}.
Moreover, in the supergravity approximation, correlation functions of the operators are 
determined by the dependence of the supergravity action on
asymptotic behaviour of the fields on the boundary. For example, the conformal
dimensions of the operators is determined by the mass of the fields.

%Fields in the $AdS$ geometries correspond to operators in the quantum field
%theory, and the mass of the fields goes over to the conformal dimension of
%the operators. An important class of operators in the conformal field theory are called
%chiral primaries. They are annihilated by some of the supersymmetries and
%so form short representations of the supersymmetry algebra. As a consequence
%their conformal dimensions are determined by how they transform under
%the $R$-symmetry. Following through the correspondence these chiral
%primaries are dual to fields in AdS space that are obtained as Kaluza-Klein
%harmonics on the reduction on the sphere. We will return to this point later.
%
%Correlation functions of the operators, which are key
%quantities in the conformal field theories, are obtained by supergravity calculations
%with assumed boundary conditions for the fields on the boundary of $AdS$.

It remains very unclear how to prove the AdS/CFT conjecture. Nevertheless,
it has now passed an enormous number of tests. Amongst the simplest
is to compare the symmetries on the two sides. ${\mathcal N}=4$ super Yang-Mills theory has
an internal $SO(6)$ ``R-symmetry'' and is
invariant under the conformal group in four-dimensions, $SO(4,2)$.
But $SO(4,2)\times SO(6)$ are precisely the isometries of $AdS_5\times S^5$. 
Moreover, after including the supersymmetry, we find that both sides are 
invariant under the action of the supergroup $SU(2,2|4)$ whose bosonic subgroup is 
$SO(4,2)\times SO(6)$. 

Let us now return to the near horizon geometries of the membrane and fivebrane.
{}For the membrane, it is conjectured that M-theory on $AdS_4\times S^7$ with 
$N$ units of flux on the seven-sphere is equivalent to a maximally
supersymmetric conformal field theory on the conformal compactification
of three-dimensional Minkowski space, the
boundary of $AdS_4$. More precisely this conformal 
field theory is the infra-red (low-energy) limit of 
$N=8$ super-Yang-Mills theory with gauge group $SU(N)$ in three dimensions. 
It is known that this theory has an $SO(8)$ R-symmetry.
For this case, the $SO(3,2)\times SO(8)$ isometries of $AdS_4\times S^7$ 
correspond to the conformal invariance and the R-symmetry of the conformal field theory. 
After including supersymmetry we find
that both sides are invariant under the supergroup $OSp(8|4)$.

{}For the fivebrane, it is conjectured that M-theory on $AdS_7\times S^4$ with 
$N$ units of flux on the four-sphere is equivalent to a maximally
supersymmetric chiral conformal field theory on the 
conformal compactification of six-dimensional Minkowski space,
the boundary of $AdS_7$. 
This conformal field theory is still rather mysterious and the AdS/CFT
correspondence actually provides a lot of useful information about it
(assuming the correspondence is valid!).
The $SO(6,2)\times SO(5)$ isometries of $AdS_7\times S^4$ correspond to the conformal invariance
and the $SO(5)$ R-symmetry of the field theory. After including supersymmetry we find
that both sides are invariant under the supergroup $OSp(6,2|4)$.
{}For the membrane and fivebrane examples, 
when $N$ is large, the radius the $AdS$ spaces are large
and M-theory is well approximated by D=11 supergravity.

Much effort has been devoted to further understanding and generalising the AdS/CFT
correspondence. Let us briefly discuss some of the generalisations that have been pursued, 
partly to put the solutions we will construct later into some kind of context, 
and partly as a rough guide to some of the vast literature on the subject.
 
The three basic examples of the AdS/CFT correspondence relate string/M-theory
on $AdS_{d+1}\times sphere$ geometries to conformally invariant quantum field
theories in d=3,4 and 6 with maximal supersymmetry. One direction is to
find new supersymmetric solutions of supergravity theories that are the 
products, possibly warped\footnote{A warped product of two spaces with
co-ordinates $x$ and $y$ corresponds to a metric of the form 
$f(y)ds^2(x)+ds^2(y)$, for some function $f(y)$.},
of $AdS_{d+1}$ with other compact spaces, 
preserving less than maximal supersymmetry. 
Since the isometry group of $AdS_{d+1}$ is the conformal group,
this indicates that these would be dual to new superconformal
field theories in $d$ spacetime dimensions. Non-supersymmetric
solutions with $AdS$ factors are also of interest, as they could
be dual to non-supersymmetric CFTs. However, 
one has to check whether the solutions are stable at both the 
perturbative and non-perturbative level, which is very difficult in general.
By contrast, in the supersymmetric case, stability is guaranteed 
from the supersymmetry algebra. The focus has thus been on supersymmetric 
geometries with $AdS$ factors.

One class of examples, discussed in \cite{Klebanov:1998hh,Acharya:1998db,Morrison:1998cs},
is to start with the fivebrane, membrane or D3-brane 
geometry \p{fivegeom}, \p{memgeom} or \p{D3geom},
respectively, and observe that if the flat space transverse to the brane is 
replaced by a manifold with special holonomy as in table 1, and the associated 
flux left unchanged, then the resulting solution will still preserve supersymmetry but
will preserve, in general, a reduced amount. Now let the special holonomy manifold
be a cone over a base $X$, i.e. let the metric of the transverse space be
\be
dr^2+r^2 ds^2(X)~,
\ee
$X$ must be Einstein and have additional well known properties to ensure that the metric
has special holonomy. {}For example a five dimensional $X$ should be Einstein-Sasaki in
order that the six-dimensional cone is Calabi-Yau.
Apart from the special case when $X$ is the round sphere
these spaces have a conical singularity at $r=0$. To illustrate this construction
explicitly for the membrane, one replaces the eight-dimensional flat space
transverse to the membrane in \p{memgeom} with an eight-dimensional cone with special holonomy:
\bea\label{genmem}
ds^2=H^{-2/3}\left[d\xi^i d\xi^j \eta_{ij}\right]
+H^{1/3}\left[dr^2+r^2 ds^2(X)\right]~,
\eea
where $H=1+\a_2N/r^6$, as before. Clearly this can be 
interpreted as $N$ co-incident membranes sitting at the conical singularity. 
By considering the near horizon limit of \p{genmem}, $r\approx 0$, one 
finds that it is now the direct product $AdS_4\times X$ and this provides a rich class
of new AdS/CFT examples.

Another generalisation is to exploit the fact that the maximally supersymmetric 
conformal field theories can be perturbed by certain operators.
In some cases, under renormalisation group flow, these quantum theories will flow in 
the infra-red (low-energies) to new super-conformal field theories, with less
supersymmetry. It is remarkable that corresponding dual supergravity solutions can be found.
Given the dictionary between operators in the conformal field theory in $d$ dimensions
and fields in $AdS_{d+1}$ mentioned above, the perturbation of the conformal field
theory should correspond to dual supergravity solutions that asymptotically tend to $AdS_{d+1}$ 
in a prescribed way. Now on rather general grounds it can be argued that this
$AdS_{d+1}$ boundary corresponds to the ultra-violet (UV) of the dual perturbed conformal field
theory and that going away from the boundary into the interior,
corresponds to going to the infra-red (IR) in the dual
quantum field theory \cite{Susskind:1998dq}. {}This can be seen, for example, 
by studying the action of  the conformal group on $AdS_{d+1}$ and on the correlation functions in the
$d$-dimensional conformal field theory.
Thus if the perturbed conformal field theory is flowing to
another conformal field theory in the IR, we expect that there should be supergravity
solutions that interpolate from the perturbed $AdS$ boundary to another $AdS$ region
in the interior. Indeed such solutions can be found (see for example 
\cite{Khavaev:1998fb,Freedman:1999gp,Pilch:2000fu}).

The above examples concern gravity duals of superconformal field theories or flows
between superconformal field theories. Another way to generalise the correspondence is to
find supergravity solutions that are dual to non-conformal field theories. 
For example, one might study perturbed superconformal field theories that
flow in the infra-red to non-conformal phases, such as Coulomb, Higgs and confining phases. 
The corresponding dual supergravity solutions should still have an asymptotic 
$AdS_{d+1}$ boundary, describing the perturbed conformal 
field theory, but they will no longer interpolate to another $AdS_{d+1}$ region but 
to different kinds of geometry dual to the different phases. {}For an example of these
kinds of solutions see \cite{Pilch:2000ue}. Often the geometries found in the IR are
singular and further analysis is required to determine the physical
interpretation.

The generalisation we will discuss in the rest of the lectures was initiated by
Maldacena and Nunez \cite{Maldacena:2000mw}. 
The idea is to construct supergravity 
solutions describing 
branes wrapping calibrated cycles in manifolds of special holonomy,
in the near horizon limit. 
The next section will explain the background for attempting this, and in particular why
such configurations preserve supersymmetry. The subsequent 
section will describe the construction of the solutions using the technical 
tool of gauged supergravity. The D=11 solutions describing wrapped membranes and fivebranes 
and the D=10 solutions describing wrapped D3-branes provide a large
class of solutions with dual field theory interpretations. The simplest, and perhaps
most important, solutions are warped products of $AdS$ spaces, cycles with Einstein 
metrics and spheres. The presence of the $AdS$ factor for these solutions
implies that they provide a large class of new AdS/CFT examples. 
In addition, as we shall discuss, there are more complicated solutions that 
describe flows from a perturbed $AdS$ boundary, describing the UV, to both
conformal and non-conformal behaviour in the IR.

Type IIB string theory also contains NS fivebranes. In the discussion section we will 
briefly discuss how supergravity solutions describing NS fivebranes wrapped on
calibrated cycles can be used to study non-conformal field theories. 
A particularly interesting solution \cite{Maldacena:2000yy} 
(see also \cite{Chamseddine:1998mc})
gives rise to a dual quantum 
field theory that has many features of ${\mathcal N}=1$ supersymmetric Yang-Mills theory 
in four dimensions. This is a very interesting theory as it 
has many features that are similar to QCD. Two other interesting ways of 
studying ${\mathcal N}=1$ supersymmetric 
Yang-Mills theory in four dimensions can be found
in \cite{Klebanov:2000hb} and \cite{Polchinski:2000uf}.
Related constructions with ${\mathcal N}=2$ supersymmetry can be found
in \cite{Gauntlett:2001ps,Bigazzi:2001aj,Bertolini:2000dk,Polchinski:2000mx} 
(for reviews see \cite{Aharony:2002up,Bertolini:2003iv,Bigazzi:2003ui}). 
The explicit regular solutions found in \cite{Klebanov:2000hb} are examples
of a more general construction discussed in \cite{Cvetic:2000mh} (see \cite{Cvetic:2002kn} for a review).

%%%%%%%%%%%
\section{Brane worldvolumes and calibrations}
%%%%%%%%%%%
The supergravity brane solutions that were presented in section 2
describe static planar branes of finite tension and infinite extent. 
Physical intuition suggests that these branes should become dynamical 
objects if they are perturbed.  Moreover, we also expect that
branes with different topologies should exist. On the other hand it 
is extremely difficult to study these aspects
of branes purely from the supergravity point of view. 
Luckily, there are alternative descriptions of branes which can be used. 
In this section we will
describe the low-energy world-volume description of branes. Essentially, this is a
probe-approximation in which the branes are taken to be
very light and hence propagate
in a fixed background geometry with no back reaction.
We will use this description to argue that branes can wrap calibrated cycles in 
manifolds of special 
holonomy while preserving supersymmetry 
\cite{Becker:1995kb,Becker:1996ay,Gauntlett:1998vk,Gibbons:1998hm}. 
This then provides the motivation to
seek D=11 supergravity solutions that describe a large number of such wrapped branes, when
the back reaction on the geometry will be very significant. We will 
construct the solutions in the next section in the near horizon limit, which is the limit
relevant for AdS/CFT applications.

It will be useful to first review some background material 
concerning calibrations on manifolds of special holonomy, 
before turning to the brane world-volume theories.

\subsection{Calibrations}

A calibration \cite{MR85i:53058} on a Riemannian manifold $M$
is a $p$-form $\varphi$ satisfying two conditions:
\bea\label{cal}
d\varphi&=&0\nn
\varphi|_{\xi^p}&\le& Vol|_{\xi^p}, \qquad \forall \xi^p~,
\eea
where $\xi_p$ is any tangent $p$-plane, and $Vol$ is the volume
form on the cycle induced from the metric on $M$. 
A $p$-cycle $\Sigma_p$ is calibrated by $\varphi$ if it satisfies
\be
\varphi|_{\Sigma_p}=Vol|_{\Sigma_p}~.
\ee
A key feature of calibrated cycles is that they are minimal surfaces
in their homology class. The proof is very simple. Consider another
cycle $\Sigma'$ such that $\Sigma-\Sigma'$ is the boundary of a 
$p+1$-dimensional manifold $\Xi$. We then have
\be
Vol(\Sigma)=\int_\Sigma \varphi=\int_\Xi d\varphi +\int_{\Sigma'}\varphi=
\int_{\Sigma'}\varphi\le Vol(\Sigma')~.
\ee
The first equality is due to $\Sigma$ being calibrated. The second
equality uses Stoke's theorem. The remaining steps use
the closure of $\varphi$ and the second part of the definition of a calibration.

We will only be interested in calibrations that can be
constructed as bi-linears of spinors, for reasons
that will become clear soon. The general procedure for 
such a construction was first discussed in \cite{dadokharvey,harvey}.
In fact all of the special 
holonomy manifolds that we discussed earlier have such calibrations. 
We now summarise the various cases, noting that the
the spinorial construction and the closure of the calibrations 
was already presented in section 2. That the calibrations also 
satisfy the second condition in \p{cal} was shown, for almost all cases,
in \cite{MR85i:53058}; 
it is also straightforward to establish using the spinorial construction.

On $Spin(7)$-holonomy manifolds the  Cayley four-form $\Psi$ is a calibration
and the 4-cycles calibrated by $\Psi$ are called Cayley 4-cycles.

$G_2$-holonomy manifolds have two types of calibrations, $\phi$ and $*\phi$.
The former calibrates associative 3-cycles, while the latter
calibrates co-associative 4-cycles.

Calabi-Yau $n$-folds generically have two classes of calibrations. 
The first class is the K\"ahler calibrations given by 
$\frac{1}{n!}J^n$ where the wedge product is used.
These calibrate even $2n$-dimensional cycles and 
this is equivalent to the cycles being holomorphic.
The second type of calibration is the special Lagrangian (SLAG) calibration
given by the real part of the holomorphic $n$-form $e^{i\theta}\Omega$, where
the constant $\theta\in S^1$, and
these calibrate special Lagrangian $n$-cycles. Recall that for
our purposes $n=2,3,4,5$. When $n=2$, there is no real distinction between 
SLAG and K\"ahler 2-cycles
since the cycles that are K\"ahler with respect to one complex structure
are SLAG with respect to another (recall that $CY_2=HK_1$).
When $n=4$ there are also 
4-cycles that are calibrated by
$\frac{1}{2}J^2+Re(e^{i\theta}\Omega)$ -- these are in fact Cayley 4-cycles if we
view the Calabi-Yau four-fold as a special example of a $Spin(7)$-manifold.

Hyper-K\"ahler manifolds in eight dimensions are special cases of Calabi-Yau
four-folds. They thus admit K\"ahler and special Lagrangian 
calibrations with respect to each complex structure. 
They also admit Cayley calibrations as just described. In addition there
are quaternionic calibrations \cite{MR89f:53089} that calibrate quaternionic
4-cycles which are K\"ahler with respect to all three complex
structures: $Vol=\frac{1}{2}(J^1)^2=\frac{1}{2}(J^2)^2=\frac{1}{2}(J^3)^2$, 
when restricted to the cycle. {}For example,
with respect to the hyper-K\"ahler structure \p{jays},
we see that $e^{1234}$ is a quaternionic 
4-cycle\footnote{Supergravity solutions
describing fivebranes wrapping quaternionic 4-cycles in $\bbR^8$, which are 
necesarrily linear \cite{MR89f:53089}, were constructed in
\cite{Papadopoulos:1998np,Papadopoulos:1998yx}.}.
Of more interest to us will be the complex-Lagrangian ($\bbC$-Lag)
calibrations \cite{MR89f:53089} which calibrate 4-cycles that are 
K\"ahler with respect to one complex structure and special Lagrangian 
with respect to the other two: for example,
$Vol=\frac{1}{2}(J^1)^2=Re(\Omega^2)=-Re(\Omega^3)$ when restricted to
the cycle. Referring to \p{jays} and \p{omegas} we see that $e^{1256}$ is an example of
such a $\bbC$-Lag 4-cycle.

In constructing supergravity solutions describing branes 
wrapping calibrated cycles in the next section, it will be very important
to understand the structure of the normal bundle of calibrated cycles
Let us summarise some results of Mclean
\cite{MR99j:53083}. The tangent bundle
of the special holonomy manifold restricted to the cycle, splits
into the tangent bundle of the cycle plus the normal bundle
\be\label{split}
T(M)|_\Sigma=T(\Sigma)\oplus N(\Sigma)~.
\ee
In some, but not all cases, the normal bundle, $N(\Sigma)$, is intrinsic to $\Sigma$.
Given a calibrated cycle, one can also ask which normal deformation, if any, is a normal
deformation through a family of calibrated cycles.

A simple case to describe are the special Lagrangian cycles, 
where $N(\Sigma)$ is intrinsic to $\Sigma$. 
It is not difficult to show that on a special Lagrangian cycle, 
the K\"ahler form, $J$, restricted to $\Sigma$ vanishes.
Thus, for any vector field $V$ on $\Sigma$ we have that 
$J_{ij}V^j$ are the components of
a one-form on $\Sigma$ which is orthogonal to all vectors on $\Sigma$. 
In other words,
$J^i{}_{j}V^j$ defines a normal vector field. 
Hence $N(\Sigma)$ is isomorphic to $T(\Sigma)$.

In addition, the normal deformation described by the vector
$V$ is a normal deformation through the space of special Lagrangian submanifolds if and
only if the one form with components $J_{ij}V^j$ is harmonic. Thus if
$\Sigma$ is compact, the dimension of the moduli space of special 
Lagrangian manifolds
near $\Sigma$, is given by the first Betti number, 
$\beta^1(\Sigma)=dim[H^1(\Sigma,\bbR)]$.
In particular if $\beta^1=0$, then $\Sigma$ has no harmonic
one-forms and hence it is rigid as a special Lagrangian submanifold.

Next consider co-associative 4-cycles in manifolds of $G_2$ holonomy, 
for which 
$N(\Sigma)$ is also intrinsic to $\Sigma$. In fact 
$N(\Sigma)$ is isomorphic to the bundle of anti-self-dual two-forms on
$\Sigma$. A normal vector field is a deformation through a family of 
co-associative 4-cycles if and only if the corresponding anti-self-dual 
two-form is closed and hence harmonic. Thus if $\Sigma$ is compact
the dimension of  the moduli space of co-associative 4-cycles
near $\Sigma$, is given by the Betti number, 
$\beta^2_-(\Sigma)=dim[H^2_-(\Sigma,\bbR)]$.
In particular if $\beta^2_-=0$, then $\Sigma$ is rigid as a co-associative submanifold.

The normal bundle of associative 3-cycles in manifolds of $G_2$ holonomy are not
intrinsic to $\Sigma$ in general. The normal bundle is given by $S\otimes V$ where
$S$ is the spin bundle of $\Sigma$ (oriented three-manifolds are always
spin) and $V$ is a rank two $SU(2)$ bundle. In other words the normal
directions are specified by two-dimensional spinors on $\Sigma$ that carry an
additional $SU(2)$ index. A normal vector field gives a deformation
through a family of associative 3-cycles if and only if the corresponding 
twisted spinor is harmonic i.e. in the kernel of the twisted Dirac operator.
In the special case that the bundle $V$ is trivial, the spinor 
must be harmonic.
For example, if $\Sigma$ is an associative three-sphere and $V$ is trivial, 
as in the $G_2$ manifold constructed in \cite{MR90i:53055}, then it is rigid.

The deformation theory of Cayley 4-cycles in manifolds of $Spin(7)$ holonomy
has a similar flavour to the associative 3-cycles. The normal bundle
is given by $S_-\otimes V$ where
$S_-$ is the bundle of spinors of negative chirality on 
$\Sigma$ and $V$ is a rank two $SU(2)$ bundle. 
Although not all 4-cycles admit a spin structure, all Cayley
4-cycles admit such twisted spinors. 
A normal vector field gives a deformation
through a family of Cayley-cycles if and only if the corresponding 
twisted spinor is harmonic i.e. in  the kernel of the twisted Dirac operator.
In the special case that the bundle $V$ is trivial the spinor must 
be harmonic.
For example, if $\Sigma$ is a Cayley four-sphere and $V$ is trivial, as
in the $Spin(7)$ manifold constructed in \cite{MR90i:53055}, 
then it is rigid.

Finally, let us make some comments concerning the K\"ahler cycles.
These cycles reside in Calabi-Yau manifolds $M$ which have vanishing first 
Chern-Class,
$c_1[T(M)]=0$. Since $c_1[T(M)|_\Sigma]=c_1[T(\Sigma)]+c_1[N(\Sigma)]$ 
we conclude that in general 
\be\label{coneid}
c_1[N(\Sigma)]=-c_1[T(\Sigma)]~.
\ee
If one considers the special case that 
$\Sigma$ is a divisor i.e. a complex hypersurface 
(i.e. real co-dimension two), 
then $N(\Sigma)$ is intrinsic to $\Sigma$. Indeed one can show that
$N(\Sigma)\cong K(\Sigma)$ where
$K(\Sigma)$ is the canonical bundle of $\Sigma$.
%In this case the moduli space has dimension 
%$\beta^{n-1,0}=dim(H^{n-1,0}(\Sigma))$

\subsection{Membrane world-volume theory}

Let us now turn to the world-volume theory of branes beginning 
with membranes \cite{Bergshoeff:1987cm,Bergshoeff:1988qx}. We will
consider the membranes to be propagating in a fixed D=11 geometry which
is taken to be a bosonic solution to the equations of motion of
D=11 supergravity, with metric $g$ and three-form $C$. 
The bosonic dynamical fields are maps $x^\mu(\s)$
from the world-volume of the membrane, $W$, to the D=11 target space geometry.
If we let $\sigma^i$ be coordinates on $W$ with $i=0,1,2$,
and $x^\mu$ be co-ordinates on the D=11 target geometry with 
$\mu=0,1,\dots,10$, the reparametrisation invariant action is given by
\be\label{memworldvol}
S=T_2\int_W d^3\sigma\left[-det\partial_i x^\mu \partial_j x^\nu g_{\mu\nu}(x)\right]^{1/2}
+\frac{1}{3!}\e^{ijk}\partial_ix^{\mu_1} \partial_jx^{\mu_2}\partial_k
x^{\mu_3} C_{\mu_1\mu_2\mu_3}~.
\ee
The first term is just the volume element of
the pull back of the metric to the world-volume and is called the 
Nambu-Goto action. The second term arises
because the membrane carries electric four-form charge; it generalises
the coupling of an electrically charged  particle to a vector potential.
The full action also includes fermions and is invariant under supersymmetry
when the D=11 target admits Killing spinors. 
The supersymmetry of brane-world volume
theories is actually quite intricate, but luckily we will not need many of the
details. The reason is similar to the reason that we didn't need to 
discuss such details for D=11 supergravity. Once again, our interest is
bosonic solutions to the equations of motion that preserve some supersymmetry. 
For such configurations, the supersymmetry variation of the 
bosonic fields automatically vanishes, and hence one only needs to know
the supersymmetry variation of the fermions, and this will be mentioned later. 

To get some further insight, consider static bosonic D=11 backgrounds with vanishing
three-form, $C=0$,and write the metric as
\be
ds^2=-dt^2+g_{MN}dx^M dx^N~,
\ee
where $M,N=1,2,\dots, 10$. If we substitute this into \p{memworldvol} and partially fix
the reparametrisation invariance by choosing $\s^0=t$ the membrane action 
gives rise to the energy functional
\be
E=T_2\int_{W'} d^2\sigma\left[m_{ab}\right]^{1/2}~,
\ee
where $a,b=1,2$, $W'$ is the spatial part of the world-volume, and 
$m_{ab}$ is the spatial part of the induced world-volume metric
given by
\be
m_{ab}=\partial_a x^M\partial_b x^N g_{MN}~.
\ee
In other words, the energy is just given by the tension of the membrane times the
spatial area of the membrane. Now, static solutions to the equations of motion
minimise the energy functional. Thus static configurations minimise
the area of the membrane, which implies that the spatial part 
of the membrane is a minimal surface. 
This is entirely in accord with expectations: the tension of 
the membrane tends to make it shrink. It should be noted that the minimal
surfaces can be of infinite extent: the simplest example being an infinite flat
membrane in D=11 Minkowski space. Of most interest to us will 
be membranes wrapping
compact minimal surfaces.

Let us further restrict to background geometries of the form $\bbR^{1,10-d}\times M_d$ with
vanishing three-form that preserve supersymmetry. In other words $M_d$ has
special holonomy as discussed in section 2. The membrane world-volume
theory is supersymmetric with the number of supersymmetries determined
by the number of Killing spinors. Static membrane configurations 
that preserve supersymmetry wrap cycles called supersymmetric cycles.
We now argue that supersymmetric cycles are equivalent to calibrated
cycles with the associated calibration being constructed from the 
Killing spinors. 

In order that a bosonic world-volume configuration be 
supersymmetric the supersymmetry
variation of the fermions must vanish. Given the explicit supersymmetry
variations, it is simple to show this implies that \cite{Becker:1995kb}
\be\label{susyconditionwv}
(1-\Gamma)\e=0~,
\ee
where $\e$ is a D=11 Killing spinor and the matrix $\Gamma$ is given
by
\bea
\Gamma&=&\frac{1}{ \sqrt{\det m}}\Gamma^0\gamma\nn
\gamma&=&\left(\frac{1}{2!}\epsilon^{ab}\partial_{a}x^M\partial_bx^N\Gamma_{MN}\right)~,
\eea
where $[\Gamma_M,\Gamma_N]_+=2g_{MN}$. The matrix $\Gamma$ satisfies
$\Gamma^2=1$ and is hermitian $\Gamma^\dagger=\Gamma$.
We now calculate 
\be
\e^\dagger\frac{(1-\Gamma)}{2}\e
=\e^\dagger\frac{(1-\Gamma)}{2}\frac{(1-\Gamma)}{2}\e
=||\frac{(1-\Gamma)}{2}\e||^2\ge 0~.
\ee
We thus conclude that $\e^\dagger \e\ge \e^\dagger\Gamma \e$ with equality
if and only $(1-\Gamma)\e=0$ which is equivalent to the configuration
being supersymmetric. The inequality can be rewritten
\be
\sqrt{\det m}\ge \e^\dagger\Gamma^0\gamma\e=-\bar\e\gamma\e~.
\ee
Thus the two-form defined by
\be\label{calspin}
\varphi=-\frac{1}{2!}\bar\e\Gamma_{MN}\e dx^M \wedge dx^N~,
\ee
satisfies the second condition in \p{cal} required for a calibration.
One can argue that the supersymmetry algebra \cite{Gutowski:1999tu} 
implies that it is closed and hence is in fact a calibration (we will verify
this directly in a moment).
Moreover, the inequality is saturated if and only if the membrane is
wrapping a supersymmetric cycle, and we see that this is
equivalent to the cycle being calibrated by \p{calspin}.

The only two-form calibrations on special holonomy backgrounds 
are K\"ahler calibrations, and indeed $\varphi$ is in fact
equal to a K\"ahler two-form on the background.  To see this 
very explicitly and to see how much supersymmetry is preserved when
a membrane wraps a K\"ahler 2-cycle, first consider the D=11
background to be $\bbR\times CY_5$. We noted earlier that 
this background preserves two D=11 supersymmetries: in a suitable orthonormal frame,
the two covariantly constant D=11 spinors can be taken to satisfy the projections
(see, for example, the discussion in appendix B of \cite{Gauntlett:2003cy}):
\be\label{projscy5}
\Gamma^{1234}\e=\Gamma^{3456}\e=\Gamma^{5678}\e=-\Gamma^{78910}\e=-\e~.
\ee
Note that these imply that $\Gamma^{012}\e =\e$. Substituting either 
of these spinors into \p{calspin} we find that $\varphi$ is precisely
the K\"ahler calibration on $CY_5$:
\be
\varphi=J=e^{12}+e^{34}+e^{56}+e^{78}-e^{910}~.
\ee
Consider now a membrane wrapping a K\"ahler 2-cycle in $CY_5$, i.e. it's worldvolume
is $\bbR\times \Sigma$ with $\Sigma\subset CY_5$. To be concrete, consider
$Vol(\Sigma)=e^{12}|_\Sigma$. We then find that the supersymmetry condition 
\p{susyconditionwv} implies
that $\Gamma^{012}\e =\e$, which is precisely the projection
on the spinors that we saw in the supergravity solution for the membrane 
\p{m2susycond}. For this case we see that
this projection does not constrain the two supersymmetries satisfying \p{projscy5} further and
thus a membrane can wrap a K\"ahler 2-cycle in a $CY_5$ ``for free''. Clearly
if we wrapped an anti-membrane, satisfying $\Gamma^{012}\e =-\e$, there would
be no surviving supersymmetry\footnote{Note that if we change
the orientation by switching $e^{10}\to -e^{10}$, then \p{projscy5} would assume
a more symmetric form and we would find that we could wrap an anti-membrane along
$\Sigma$ for free.}. Let us now consider the background to be 
$\bbR\times CY_4\times \bbR^2$. This
preserves four supersymmetries satisfying projections which we can take to be
\be\label{projscy4}
\Gamma^{1234}\e=\Gamma^{3456}\e=\Gamma^{5678}\e=\mp\Gamma^{78910}\e=-\e~.
\ee
Two of these satisfy $\Gamma^{012}\e =\e$ and two satisfy $\Gamma^{012}\e =-\e$.
After substituting into \p{calspin} they give rise to two K\"ahler forms on 
$CY_4\times \bbR^2$:
\be
\varphi=J=e^{12}+e^{34}+e^{56}+e^{78}\mp e^{910}~.
\ee
If we now wrap the membrane on a K\"ahler 2-cycle with $Vol(\Sigma)=e^{12}|_\Sigma$, then we
see that the supersymmetry condition \p{susyconditionwv}, $\Gamma^{012}\e =\e$,
preserves two of the supersymmetries. Similarly, if we wrapped an anti-membrane
satisfying $\Gamma^{012}\e =-\e$ it would also preserve two supersymmetries.

The amount of supersymmetry preserved by any brane wrapping a calibrated cycle 
in a special holonomy background can be worked out in a similar way: 
one considers a convenient set of projections for the background geometry and
then supplements them with those of the wrapped brane (or anti-brane). In almost
all cases, wrapping the brane breaks 1/2 of the supersymmetries preserved
by the special holonomy background. We have summarised the possibilities 
for the membrane in table 2.

\begin{table}[htbp]
\begin{center}
\begin{tabular}{|c|c|c|}
   \hline
   Calibration& World-Volume& Supersymmetry \\ 
   \hline
   \hline
   K\"ahler& $\bbR\times (\Sigma_2\subset CY_2)$& 8\\ 
   \cline{2-3}
           & $\bbR\times (\Sigma_2\subset CY_3)$& 4 \\
   \cline{2-3}
           & $\bbR\times (\Sigma_2\subset CY_4)$& 2 \\
   \cline{2-3}
           & $\bbR\times (\Sigma_2\subset CY_5)$& 2 \\
   \hline
\end{tabular}
   \caption{The different ways in which membranes can wrap calibrated cycles 
     and the amount of supersymmetry preserved.}
   \label{tab:tablemem}
\end{center}
\end{table}

The action \p{memworldvol} describes the dynamics of a membrane propagating in
a fixed D=11 supergravity background. Such a membrane is often called a ``probe
membrane''.
Of course, the dynamics of the membrane will back react on the geometry,
and so one should really supplement the D=11 supergravity action with the world-volume
action:
\be\label{combinedact}
S=S_{D=11}+S_{WV}~.
\ee
If there are many coincident membranes then
this back reaction could be large.

We have been emphasising the geometric aspects of the membrane world-volume
theory. The world-volume theory is also a quantum field theory. To gain some insight
into this aspect, let us restrict the target geometry to be D=11 Minkowski
space and the world-volume to be $\bbR^{1,2}$. 
Now fix the reparametrisation invariance completely by
setting $\sigma^0=t,\s^1=x^1,\s^2=x^2$. We can then expand the determinant to
get
\be\label{gfixed}
S=T_2\int d^3\sigma (-\frac{1}{2}\partial_a x^I \partial^a x^I+fermions+\dots)~,
\ee
where we have dropped an infinite constant and the dots refer to higher
derivative terms. The eight scalar fields describe the eight transverse 
fluctuations of the membrane.
After quantisation, this action gives a three-dimensional quantum field
theory, with eight scalar fields plus fermions, that 
preserves 16 supersymmetries or $N=8$ supersymmetry in three dimensions.
This quantum field theory is interacting with gravity, via \p{combinedact},
but if we take the limit, $l_{p}\to 0$, it decouples from gravity. In other words, 
in this decoupling limit we get a three-dimensional quantum field theory living on
the world-volume of the membrane. When there are $N$ coincident branes, the
quantum field theory is much more complicated. There is a piece describing the
centre of mass dynamics of the branes given by \p{gfixed} with $T_2\to NT_2$
and there is another piece describing the interactions between the membranes.
This latter theory is known to be a superconformal field theory that arises
as the IR limit of $N=8$ supersymmetric Yang-Mills theory in three dimensions.
Recall that this is precisely the superconformal field theory that 
is conjectured to be dual to M-theory on $AdS_4\times S^7$.

The important message here is that the supergravity solution describing the membranes in
the near horizon limit,  $AdS_4\times S^7$, is conjectured to be equivalent to the quantum
field theory arising on the membrane world-volume 
theory, in a limit which decouples gravity. 

Now consider a more complicated example. Take the 
D=11 background to be of the form $\bbR^{1,6}\times CY_2$ with
a probe membrane wrapping a  K\"ahler 2-cycle $\Sigma\subset CY_2$. i.e.
the world-volume of the membrane is $\bbR\times (\Sigma\subset CY_2)$.
There is again a quantum field theory living on the brane interacting with gravity.
In the decoupling limit, $l_{p}\to 0$ and keeping the volume of $\Sigma$ fixed, 
we get a supersymmetric quantum field theory on $\bbR\times \Sigma$. 
If $\Sigma$ is compact, the low-energy infra-red (IR) limit of this quantum field theory,
corresponds to length scales much larger than the size of $\Sigma$. In this IR limit
the quantum field theory on $\bbR\times \Sigma$ behaves like a quantum field theory 
on the time direction $\bbR$, which is just a quantum mechanical model.

If we could construct a supergravity solution describing
membranes wrapping such K\"ahler 2-cycles, in the near horizon limit, we 
would have an excellent candidate for an M-theory dual for this quantum field
theory on $\bbR\times \Sigma$. Moreover, if the supergravity solution has an 
$AdS_2$ factor, it would strongly indicate that the corresponding 
dual quantum mechanics, arising in the IR limit, is a superconformal quantum mechanics.
These kind of supergravity solutions have been found \cite{Gauntlett:2001qs} 
and the construction will be described 
in the next section.

It is worth making some further comments
about the quantum field theory on $\bbR\times \Sigma$. For a single membrane the physical
bosonic degrees of freedom describe the transverse deformations of the
membrane. In the case of a membrane with world-volume $\bbR^{1,2}$
in $\bbR^{1,10}$ we saw above in \p{gfixed} 
that there are eight scalar fields 
describing these deformations. Geometrically, they are sections
of the normal bundle, which is trivial in this case. 
Now consider, for example, a membrane with world-volume $\bbR\times (\Sigma \subset CY_2)$. 
There are six directions transverse to the membrane that are also transverse to the $CY_2$
and these lead to six scalar fields. There are also two directions
transverse to the membrane that are tangent to the $CY_2$: these give
rise to a section of the normal bundle. As we discussed earlier the normal
deformations of a K\"ahler 2-cycle $\Sigma \subset CY_2$ (which are also SLAG 2-cycles
with respect to another complex structure) are specified by one-forms on $\Sigma$. 

This ``transition'' from scalars to one-forms is intimately connected with the
way in which the field theory on $\bbR\times \Sigma$ realises supersymmetry.
In particular it arises because the theory is coupled to external 
R-symmetry gauge-fields. We will discuss this issue again in the context of 
fivebranes wrapping SLAG 3-cycles, as this is the example we will focus on 
when we construct the supergravity solutions in the next section.

\subsection{D3-brane and fivebrane world-volume theories}

Let us now briefly discuss the world-volume theories of the type IIB
D3-brane and the M-theory fivebrane. The D3-brane 
action is given by a Dirac-Born-Infeld type action that includes a
coupling to a four-form potential whose field strength is the 
self-dual five-form (see Myers' lectures for further discussion). 
If we consider, for simplicity, a bosonic type IIB background with 
all fields vanishing except for the metric, the world-volume 
action for a single D3-brane, with fermions set to zero, is given
by
\be\label{d3worldvol}
S=T_3\int_W d^4\sigma
\left[-det(\partial_i x^\mu \partial_j x^\nu g_{\mu\nu}(x)+F_{ij})\right]^{1/2}~.
\ee
The main new feature is that, in addition to the 
world-volume fields $x^\mu$, there
is now a $U(1)$ gauge-field with field strength $F$. 

If we set $F=0$, the action \p{d3worldvol} reduces to the Nambu-Goto action. Following
a similar analysis to that of membrane, we again
find in static supersymmetric backgrounds of the form $\bbR^{1,10-d}\times M_d$ that
supersymmetric cycles with $F=0$ are calibrated cycles. As the D3-brane has three spatial
world-volume directions, there are now more possibilities. A D3-brane can either
wrap a calibrated 3-cycle in $M_d$, with world-volume 
$\bbR\times(\Sigma_3\subset M_d)$ or a K\"ahler 2-cycle in 
$CY_n$ with world-volume $\bbR^{1,1}\times(\Sigma_2\subset CY_n)$. The possibilities
with the amount of supersymmetry preserved are presented in table~\ref{tab:tabled3}. {}For
the K\"ahler cases, the world-volume has an $\bbR^{1,1}$ factor and 
we have also denoted by $(n_+,n_-)$ the amount $d=2$ supersymmetry preserved
on the $\bbR^{1,1}$ factor, where $n_+$ is the number of chiral supersymmetries
and $n_-$ the number of anti-chiral supersymmetries.

\begin{table}[htbp]
\begin{center}
\begin{tabular}{|c|c|c|}
   \hline
   Calibration& World-Volume& Supersymmetry \\ 
   \hline
   \hline
   K\"ahler& $\bbR^{1,1}\times (\Sigma_2\subset CY_2)$&8,\quad  (4,4) d=2\\ 
   \cline{2-3}
           & $\bbR^{1,1}\times (\Sigma_2\subset CY_3)$&4,\quad (2,2) d=2 \\
   \cline{2-3}
           & $\bbR^{1,1}\times (\Sigma_2\subset CY_4)$&2,\quad (1,1) d=2\\
   \hline
   SLAG   & $\bbR\times (\Sigma_3\subset CY_3)$ & 4 \\ 
   \cline{2-3}
   \hline
   Associative  & $\bbR\times (\Sigma_3\subset G_2)$ & 2 \\ 
   \cline{2-3}
   \hline
\end{tabular}
   \caption{The different ways in which D3-branes can wrap 
     calibrated cycles and the amount of supersymmetry preserved.}
   \label{tab:tabled3}
\end{center}
\end{table}

In order to get some insight into the field theory living on D3-branes, consider
the target to be D=10 Minkowski space-time, the world-volume to be $\bbR^{1,3}$ and 
fix the reparametrisation invariance by setting $\sigma^0=t,\s^1=x^1,\s^2=x^2,\s^3=x^3$. 
After expanding the determinant and dropping a constant term, we get 
\be\label{d3gfixed}
S=T_3\int d^4\sigma (-\frac{1}{2}\partial_i x^I \partial^i x^I-\frac{1}{4}F_{ij}F^{ij}+fermions+\dots)~.
\ee
In addition to $F$ there are six scalar fields that describe the transverse
displacement of the D3-brane. This action
is simply $N=4$ super-Yang-Mills theory with gauge group $U(1)$. When there are
$N$ D3-branes, it is known that the DBI action should be replaced by a non-abelian
generalisation but its precise form is not yet known. However, it is known that after
decoupling gravity, the
leading terms give $U(N)$ ${\mathcal N}=4$ super-Yang-Mills theory. After
dropping the $U(1)$ centre of mass piece, we find ${\mathcal N}=4$ $SU(N)$ Yang-Mills theory.
Recall that this is the theory that is conjectured to be dual to type 
IIB string theory on $AdS_5\times S^5$, which is the 
near horizon limit of the type IIB supergravity solution describing a planar D3-brane.
Once again, the near horizon limit of the supergravity solution
is dual to the field theory arising on the brane, and the same should apply
to the near horizon limits of the supergravity solutions describing the wrapped D3-branes in
table~\ref{tab:tabled3}.

The world-volume theory of M-theory fivebranes is arguably the most intricate of all
branes \cite{Howe:1997yn,Howe:1997fb,Pasti:1997gx,Bandos:1997ui,Bergshoeff:1998vx}. 
The bosonic dynamical fields are maps $x^\mu(\sigma)$ along with a three-form
field strength $H_{ijk}$ that satisfies a non-linear self-duality condition.
If we set $H=0$ the dynamics is described by the Nambu-Goto action and
we again find in a static supersymmetric background of the form $\bbR^{1,10-d}\times M_d$ that
supersymmetric cycles with $H=0$ are calibrated cycles. As the fivebrane has five
spatial world-volume directions, there are many possibilities
which are summarised in table~\ref{tab:tablefive}. We have included the amount
of supersymmetry preserved including the number of supersymmetries
counted with respect to the flat part of the world-volume
$\bbR^{1,q}$ when $q\ge 1$.

\begin{table}[htbp]
\begin{center}
\begin{tabular}{|c|c|c|}
   \hline
   Calibration& World-Volume& Supersymmetry \\ 
   \hline
   \hline
   SLAG& $\bbR^{1,3}\times (\Sigma_2\subset CY_2)$& 8,\quad${\mathcal N}$=2 d=4\\ 
   \cline{2-3}
      & $\bbR^{1,2}\times (\Sigma_3\subset CY_3)$& 4,\quad${\mathcal N}$=2 d=3\\ 
   \cline{2-3}
      & $\bbR^{1,1}\times (\Sigma_4\subset CY_4)$& 2,\quad(1,1) d=2\\ 
   \cline{2-3}
      & $\bbR\times (\Sigma_5\subset CY_5)$& 1\\ 
 \cline{2-3}
      & $\bbR^{1,1}\times (\Sigma_2\subset CY_2)\times(\Sigma_2'\subset CY_2')$& 4,
\quad(2,2) d=2\\ 
 \cline{2-3}
      & $\bbR\times (\Sigma_2\subset CY_2)\times (\Sigma_3\subset CY_3)$& 2\\ 
\hline
   K\"ahler& $\bbR^{1,3}\times (\Sigma_2\subset CY_3)$ & 4,\quad${\mathcal N}$=1 d=4\  \\ 
   \cline{2-3}
         & $\bbR^{1,1}\times (\Sigma_4\subset CY_3)$ & 4,\quad (4,0) d=2\  \\ 
   \cline{2-3}
         & $\bbR^{1,1}\times (\Sigma_4\subset CY_4)$ & 2,\quad (2,0) d=2\  \\ 
   \hline
   $\bbC$-Lag& $\bbR^{1,1}\times (\Sigma_4\subset HK_2)$ & 3,\quad (2,1) d=2\  \\ 
\hline
   Associative  & $\bbR^{1,2}\times (\Sigma_3\subset G_2)$ & 2,\quad${\mathcal N}$=1 d=3\\
   \hline
   Co-associative  & $\bbR^{1,1}\times (\Sigma_4\subset G_2)$ & 2,\quad (2,0) d=2\\ 
   \hline
   Cayley  & $\bbR^{1,1}\times (\Sigma_4\subset Spin(7))$ & 1,\quad (1,0) d=2 \\ 
   \hline
\end{tabular}
   \caption{The different ways in which fivebranes can wrap calibrated cycles 
   and the amount of
   supersymmetry preserved.}
   \label{tab:tablefive}
\end{center}
\end{table}

The six-dimensional 
field theory living on a single planar fivebrane has five scalar fields, describing the 
transverse fluctuations, the three-form $H$ and fermions, 
and has a chiral $(2,0)$ supersymmetry. The field theory
when there are $N$ coincident fivebranes is not yet well understood. The AdS/CFT
conjecture states that it is dual to M-theory propagating on $AdS_7\times S^4$.
Recall that the field theory has an $SO(5)$ R-symmetry.

In oder to construct supergravity solutions describing wrapped
branes, it is very helpful to understand how supersymmetry is realised 
in the field theory living on a probe-brane world-volume. 
The details depend on which calibrated cycle is
being wrapped and it is intimately connected to the structure of the normal
bundle of the calibrated cycle. Let us concentrate on the case of fivebranes
wrapping SLAG 3-cycles, as this will be the focus of the next section.
The field theory on the probe fivebrane world-volume lives on $\bbR^{1,2}\times \Sigma_3$.
In order for this field theory to be supersymmetric it is necessary that there
is some notion of a constant spinor on $\Sigma_3$. It is not immediately clear
what this notion is, since, in general, $\Sigma_3$ will not have a covariantly 
constant spinor. However, the field theory living on the fivebrane with world-volume
$\bbR^{1,5}\subset \bbR^{1,10}$ has an internal $SO(5)$ R-symmetry, coming from the
five flat transverse directions, under which the fermions transform.
The covariant derivative of the spinors is schematically of the form
\be\label{twistee}
(\partial_\mu+\omega_\mu-A_\mu)\e~,
\ee
where $\omega$ is the spin connection and $A$ is the $SO(5)$ gauge connection.
Now consider the fivebrane theory on $\bbR^{1,2}\times \Sigma_3$.
If we decompose $SO(5)\to SO(3)\times SO(2)$  and choose the $SO(3)$
gauge-fields to be given by the $SO(3)$ spin connection on $\Sigma_3$, $A=\omega$,
then clearly we can have constant spinors on $\Sigma_3$ that could
parametrise the supersymmetry. This is exactly the way
supersymmetry is realised for wrapped branes \cite{Bershadsky:1996qy}. 
It is sometimes said that the field theory is ``twisted'' because of the similarities with
the construction of topological field theories.

Geometrically, the identification of the $SO(3)\subset SO(5)$ gauge fields
with the spin connection on $\Sigma_3$ corresponds to the structure of normal bundle
of a SLAG 3-cycle. The five directions transverse to the fivebrane wrapping
the SLAG 3-cycle consist of three directions that are tangent to the 
$CY_3$ and two flat directions that
are normal to the $CY_3$. This is responsible for breaking the $SO(5)$ symmetry 
of the flat fivebrane down to $SO(3)\times SO(2)$. 
We expect an $SO(2)$ R-symmetry to survive 
corresponding to the two flat directions, and thus the $SO(2)$ gauge
fields are zero in the vacuum state. In section 4.1 we argued that
for SLAG 3-cycles $N(\Sigma_3)\cong T(\Sigma_3)$ and this is responsible
for the fact that there are non-zero $SO(3)$ gauge-fields in the vacuum state and
moreover, $A=\omega$. 

We can determine which external $SO(5)$ gauge-fields are excited for
fivebranes wrapping different supersymmetric cycles from our previous discussion
of the normal bundles of calibrated cycles. We shall mention this again
in the next section in the context of constructing the corresponding
supergravity solutions.

\subsection{Generalised Calibrations}
As somewhat of an aside, we comment that there are more general supersymmetric 
cycles than those we have discussed above.

{}Firstly, we only considered background geometries when 
the background fluxes (e.g. the four-form field strength $G$ for D=11 supergravity) 
are all set to zero. 
By analysing supersymmetric brane configurations when
the fluxes are non-zero, one is naturally lead to 
the notion of ``generalised calibrations'' \cite{Gutowski:1999iu,Gutowski:1999tu} 
(see also \cite{Barwald:1999ux}).
The key new feature is that the exterior derivative of the generalised 
calibration is now longer zero and is related to the flux.  
It is interesting to note that generalised calibrations play an important role
in characterising the most general classes of supersymmetric
supergravity solutions 
\cite{Gauntlett:2001ur,Gauntlett:2002sc,Gauntlett:2002fz,Gauntlett:2003cy}.
They have also been discussed in \cite{Gutowski:2002bc}.

A second generalisation, is to determine the conditions
for supersymmetric branes when non-trivial world-volume fluxes ($F$ for D3-branes
and $H$ for the fivebranes) are switched on. This is related to the possibility
of branes ending on branes and is discussed in 
\cite{Bergshoeff:1997kr,Gauntlett:1998wb,Gauntlett:1999aw,Marino:1999af}.

\section{Supergravity solutions for fivebranes wrapping calibrated cycles}

At this point, we have established that D=11 supergravity has 
supersymmetric membrane and fivebrane solutions. 
By considering the world-volume approximation to the dynamics of these branes we
concluded that supersymmetric solutions of
D=11 supergravity describing branes wrapping calibrated cycles
in special holonomy manifolds should also exist.
In this section we will explain the explicit construction of such solutions,
in the near horizon limit, focusing on the richest case of fivebranes.

At first sight it is not at all clear how to construct these solutions.
One might imagine that one should start with an explicit special holonomy
metric, which are rather rare, and then ``switch on the brane''. In fact
the procedure we adopt \cite{Maldacena:2000mw} 
is more indirect and subtle. A key point is that we aim
to find the solutions in the near horizon limit, i.e. near to the brane
wrapping the cycle, and this simplifies things in two important ways. 
Firstly, we expect that only the local geometry of the calibrated cycle
in the special holonomy manifold,
including the structure of its normal bundle, to enter into
the construction.  Secondly, we will be able to employ a very useful
technical procedure of first finding the solutions in
D=7 $SO(5)$ gauged supergravity. This theory arises from the consistent truncation
of the  
Kaluza-Klein reduction on a four-sphere of D=11 supergravity, as we shall describe. 
In particular any supersymmetric solution of the
D=7 theory gives rise to a supersymmetric solution of D=11 supergravity.
Although, the converse is certainly not true, the D=7 gauged supergravity
does include many interesting solutions corresponding to the near horizon
limit of wrapped fivebrane geometries.

We first discuss Kaluza-Klein reduction starting with the simplest case
of reduction on a circle. We then describe the reduction on a four-sphere
leading to D=7 gauged supergravity. {}Following this we will describe the construction
of fivebranes wrapping calibrated cycles. We focus 
on the case of fivebranes wrapping SLAG 3-cycles to illustrate some details
and then summarise some aspects of the other cases.

\subsection{Consistency of Kaluza-Klein reduction}

The basic example of Kaluza-Klein dimensional reduction is to start with pure
gravity in five spacetime dimensions and then reduce on a circle to get
a theory of gravity in four spacetime dimensions coupled to a $U(1)$ gauge
field and a scalar field. The procedure is to first expand the five-dimensional metric
in harmonics on the circle. One obtains an infinite tower of modes
whose four-dimensional mass is proportional to the inverse of
the radius of the circle, as well as some massless modes consisting of the
four dimensional metric, gauge field and scalar field just mentioned.
Finally one truncates the theory to the massless mode sector. 

This truncation is said to be ``consistent'' 
in the sense that any solution of the four-dimensional theory 
is automatically a solution to the five-dimensional
theory. The reason for this consistency is simply that the massless modes 
being kept are independent of the coordinate on the circle, 
while the massive modes, which have
non-trivial dependence on the coordinate on the circle, are all set to zero. 
Note that this is an exact statement that does not rely
on the radius of the circle being small, where one might argue that the massive
modes are decoupling because they are all getting very heavy. 
Similarly, as we shall shortly illustrate in more detail, one can consistently 
truncate the reduction of theories with additional matter fields on a circle, 
and more generally on tori. 

The D=7 gauged supergravity that we shall be interested in arises from the 
dimensional reduction of D=11 supergravity on a four-sphere. In general
there are no consistent truncations of the dimensional reductions of gravity 
theories on spheres with dimension greater than one. 
The reason is that all of the harmonics on the sphere,
including those associated with the lowest mass modes, typically depend on 
the coordinates of the sphere. Indeed, generically, if one reduces a theory of 
gravity on a sphere and attempts to truncate to the lowest mass modes, 
one will find that it is not consistent. That is, solutions of the truncated theory 
will not correspond to exact solutions of the higher dimensional theory.
Of course if the radius of the sphere were taken to be very small the truncated
solutions could provide very good approximations to solutions of the higher dimensional
theory. However, in some special cases, including the reduction of 
D=11 supergravity on a four-sphere, it has been shown that
there is in fact a consistent truncation. We will exploit this fact to construct exact
solutions of D=11 supergravity by ``uplifting'' solutions that we first find
in the D=7 gauged supergravity.

Before we present the Kaluza-Klein reduction formula for D=7 gauged supergravity,
which are rather involved, let us first present them in the much simpler setting of 
type IIA supergravity.

\subsection{Reduction on $S^1$ to type IIA supergravity }
Type IIA supergravity in ten dimensions can be obtained from the Kaluza-Klein
reduction of D=11 supergravity on $S^1$ \cite{Giani:1984wc,Campbell:1984zc,Huq:1985im}. 
To see this, we construct an
ansatz for the D=11 supergravity fields that just maintains the lowest
massless modes. For the bosonic fields we let
\bea\label{iiared}
ds^2&=&e^{-2\Phi/3}ds^2_{10}+e^{4\Phi/3}(dy+C^{(1)})^2\nn
C&=& C^{(3)} +B\wedge dy~.
\eea
Here the ten dimensional line element $ds^2_{10}$, the scalar dilaton $\Phi$,
the ``Ramond-Ramond'' one-form $C^{(1)}$ and three-form $C^{(3)}$, and
the the Neveu-Schwarz two-from $B$ are all independent of $y$.
The field strengths of the forms will be denoted $F^{(2)}=dC^{(1)}$,
$F^{(4)}=dC^{(3)}$ and $H=dB$. If we substitute this ansatz into the D=11 equations
of motion we find equations of motion for the ten-dimensional fields
which are derivable from the action 
\be
S=\int d^{10} x {\sqrt -g}\left(e^{-2\Phi}[R+4\partial\Phi^2-\frac{1}{12} H^2]
-\frac{1}{48}F_{(4)}^2-\frac{1}{4}F_{(2)}^2\right)-\frac{1}{2}B\wedge F_4\wedge F_4~.
\ee
This is precisely the bosonic part of the action of type IIA
supergravity. After similarly including the fermions we find
the full supersymmetric type IIA action, which preserves 32 supersymmetries
(two D=10 Majorana-Weyl spinors of opposite chirality). 
Note that the isometries of $S^1$ give rise to
the $U(1)$ gauge field with field strength
$F_{(2)}$.

The key point to emphasise is that, by construction, any solution of
type II supergravity automatically can be uplifted to give a solution of D=11
supergravity that admits a $U(1)$ isometry using the formulae in \p{iiared}. 
Moreover, the D=11 supergravity solution will preserve at least the same amount 
of supersymmetry as the type IIA solution\footnote{Note that there 
are D=11 solutions with $U(1)$ isometries that have supersymmetries that will not 
survive the dimensional reduction because the Killing spinors have a non-trivial 
dependence on the coordinate on the circle.}.

To illustrate with a simple example, consider the
following supersymmetric solution of type IIA supergravity:
\bea
ds^2&=&H^{-1}(d\xi^i d\xi^j\eta_{ij})+(dx^Idx^I)\nn
B&=&H^{-1}d\xi^0\wedge d\xi^1\nn
e^{2\Phi}&=&H^{-1}~,
\eea
with $i,j=0,1$ and $I,J=1,\dots, 8$.
This is a solution to IIA supergravity providing that $H$ is harmonic
in the transverse space. Choosing the simple single centre solution
$H=1+\a_2N/r^6$ we find that this solution carries $N$ units of
quantised electric $H$-flux. In fact this solution describes the fields around $N$ co-incident
fundamental IIA strings of infinite extent. 
It preserves one-half of the supersymmetry. If we now uplift this solution to get a solution
of D=11 supergravity using \p{iiared} we obtain the planar membrane solution \p{memgeom}.

\subsection{Reduction on $S^4$ to D=7 $SO(5)$ gauged supergravity}

The dimensional reduction of D=11 supergravity 
on a four-sphere can be consistently truncated to give D=7 
$SO(5)$ gauged supergravity \cite{Nastase:1999cb,Nastase:1999kf}. The origin of the 
$SO(5)$ gauge symmetry is the $SO(5)$ isometries of the four-sphere.

The explicit formulae for the D=11 bosonic fields is given by
\bea\label{d7redform}
ds^2&=&\Delta^{-2/5}ds^2_7+\frac{\Delta^{4/5}}{m^2}DY^A (T^{-1})^{AB} DY^B\nn
8G&=&\epsilon_{A_1...A_5}\left[-\frac{1}{3m^3}
DY^{A_1}DY^{A_2}DY^{A_3}DY^{A_4}\frac{(T\cdot Y)^{A_5}}{Y\cdot T\cdot Y}
\right.\nonumber\\&&\left. +\frac{4}{3m^3}
DY^{A_1}DY^{A_2}DY^{A_3}D\left(\frac{(T\cdot Y)^{A_4}}{
Y\cdot T\cdot Y}\right)Y^{A_5}\right.\nonumber\\
&&\left.+\frac{2}{m^2}F^{A_1A_2}DY^{A_3}DY^{A_4}\frac{(T\cdot
Y)^{A_5}}{Y\cdot T\cdot Y}+\frac{1}{m}
F^{A_1A_2}F^{A_3A_4}Y^{A_5}\right]\nonumber\\&&
+d(S_{B}Y^B)~,
\eea   
and the wedge product of forms is to be understood in the expression for $G$.
Here $Y^A$, $A,B=1,\dots ,5$, are constrained co-ordinates
parametrising a four-sphere, satisfying $Y^A Y^A=1$. In this section
$x^\mu$, $\mu,\nu=0,1,\dots,6$ are D=7 co-ordinates and
$ds^2_7$ is the D=7 line element associated with the D=7 metric
$g_7(x)$. The $SO(5)$ isometries of the round four-sphere lead to the
introduction of $SO(5)$ gauge-fields $B^A{}_B(x)$, with field strength 
$F^A{}_B(x)$,
that appear in the covariant derivative $DY^A$:
\be
DY^A=dY^A+2m B^A{}_B Y^B~.
\ee
The matrix $T$ is defined by
\be
T^{AB}(x)=(\Pi^{-1})_i{}^A(x) (\Pi^{-1})_j{}^B(x) \delta^{ij}~,
\ee
where $i,j=1,\dots,5$ and $\Pi_A{}^i(x)$ is 14 scalar fields that parametrise the 
coset $SL(5,R)/SO(5)$. The warp factor $\Delta$ is defined via
\be
\Delta^{-6/5}=Y^A T^{AB}(x)Y^B
\ee
and $S_A(x)$ are five three-forms.

If we substitute this into the D=11 supergravity equations of motion we
get equations of motion for $g_7(x) ,B(x),\Pi(x)$ and $S(x)$. The resulting equations of motion
can be derived from the D=7 action: 
\begin{equation}
\begin{aligned}
  S&= \int d^7x\sqrt{-g} \bigg[ R
         + \frac{1}{2}m^2\left(T^2-2T_{ij}T^{ij}\right)
         - P_{\mu ij}P^{\mu ij}
         - \frac{1}{2} \left({\Pi_A}^i{\Pi_B}^jF_{\mu\nu}^{AB}\right)^2
      \\ & \qquad\qquad\qquad\qquad
         - m^2 \left({{\Pi^{-1}}_i}^AS_{\mu\nu\rho,A} \right)^2
      \bigg]
      - 6m S_A\wedge F_A
      \\ & \qquad\qquad
      + \sqrt{3}\epsilon_{ABCDE}
          \delta^{AG}S_G\wedge F^{BC}\wedge F^{DE}
      + \frac{1}{8m} \left(2\Omega_5[B]-\Omega_3[B]\right)~,
\end{aligned}
\label{Ssugra}
\end{equation}
where have dropped the 7 subscript on $g_7$ here and below, for clarity.
This is precisely the bosonic part of the action of D=7 $SO(5)$ gauged supergravity 
\cite{Pernici:1984xx}.
In particular, the kinetic energy terms for the scalar fields are determined by
$P_{\mu ij}$ which is defined to be the part symmetric in $i$ and $j$ of 
\be\label{psandqs}
{(\Pi^{-1})_i}^A\left({\delta_A}^B \partial_\mu + 
2m B_{\mu\,A}{}^B\right){\Pi_B}^k \delta_{kj}~.
\ee
The potential terms for the scalar fields are defined in terms
of 
\be
T_{ij}=(\Pi^{-1})_i{}^A(\Pi^{-1})_j{}^A,\qquad T=T_{ii}~.
\ee
Finally, $\Omega_3[B]$ and $\Omega_5[B]$ are Chern-Simons forms for the gauge-fields,
whose explicit form will not be needed.

It is also possible to explicitly construct an ansatz for the D=11 fermions to 
recover the fermions and supersymmetry of the D=7 gauged supergravity.
In particular the construction implies 
that any bosonic (supersymmetric) solution of D=7 $SO(5)$ gauged supergravity
will uplift via \p{d7redform} to a bosonic (supersymmetric) solution of D=11 supergravity. 

In order to find bosonic supersymmetric solutions of the D=7 gauged supergravity,
we need the supersymmetry variations of the fermions. The fermions consist of
gravitini $\psi_\mu$ and dilatini  $\lambda$ and, in a bosonic background, 
their supersymmetry variations are given by
\bea\label{susytran}
\delta\psi_{\mu}&=&\nabla_{\mu}\epsilon
 -\frac{1}{40}(\gamma_{\mu}\;^{\nu\rho}-8\delta_{\mu}^{\nu}\gamma^{\rho})\Gamma_{ij}\epsilon\, {\Pi_A}^i{\Pi_B}^j
F_{\nu\rho}^{AB}
\nonumber\\
&& +\frac{1}{20}mT\gamma_{\mu}\epsilon+ \frac{m}{10\sqrt{3}}(\gamma_{\mu}\;^{\nu\rho\sigma}
-\frac{9}{2}\delta_{\mu}\;^{\nu}\gamma^{\rho\sigma})\Gamma^i\epsilon\,
{{\Pi^{-1}}_i}^AS_{\nu\rho\sigma,A} \nn
\delta\lambda_i&=&         
 \frac{1}{2}\gamma^{\mu}\Gamma^j \epsilon\, P_{\mu ij}
   + \frac{1}{16}\gamma^{\mu\nu}(\Gamma_{kl}\Gamma_i
       -\frac{1}{5}\Gamma_i\Gamma_{kl})
       \epsilon\, {\Pi_A}^k{\Pi_B}^l F_{\mu\nu}^{AB} \nonumber\\
   &&    + \frac{1}{2}m(T_{ij}-\frac{1}{5}T\delta_{ij})\Gamma^j\epsilon
+ \frac{m}{20\sqrt{3}}\gamma^{\mu\nu\rho}(\Gamma_i\; ^j-4\delta_i^j)
      \epsilon\,{{\Pi^{-1}}_j}^A S_{\mu\nu\rho ,A}~.
\eea
Here $\gamma^\mu$ are the D=7 gamma matrices of $Cliff(6,1)$, while $\Gamma^i$ are those
of $Cliff(5)$, and these act on $\e$ which is a spinor with respect to both $Spin(6,1)$ and
$Spin(5)$. Note that $\gamma^\mu$ and $\Gamma^i$ commute. The covariant derivative 
appearing in the supersymmetry variation for the gravitini is given by
\be\label{nabla}
   \nabla_\mu\epsilon = \left( \partial_\mu
       + \frac{1}{4}\omega_\mu^{ab}\gamma_{ab}
       + \frac{1}{4}Q_{\mu ij}\Gamma^{ij} \right) \epsilon~,
\ee   
where $Q_{\mu ij}$ is the part of \p{psandqs} anti-symmetric in $i$ and $j$. To obtain
supersymmetric configurations we set $\delta\psi=\delta\lambda^i=0$ and let
$\e$ be a commuting spinor. Note that setting the variations of the dilatini
to zero leads to algebraic constraints on $\e$.

Although still very complicated, D=7 gauged supergravity is a simpler
theory than D=11 supergravity as many degrees of freedom have been truncated.
It is clear from \p{d7redform} that the ``breathing mode'', 
corresponding to uniformly scaling the
round four-sphere, is one of the modes that has been truncated. This means
that there are no solutions of the D=7 gauged supergravity which give rise
to the full fivebrane solution of D=11 supergravity \p{fivegeom} with
\p{m5singcent}. However, the
near horizon limit of the fivebrane solution, $AdS_7\times S^4$, is easily
found. Indeed it arises as the simplest ``vacuum'' solution of the 
theory where the gauge fields, the three-forms and the scalars are all set to
zero: $B=S=0$, $\Pi=\delta$. In this case the equations of motion
reduce to solving
\be
R_{\mu\nu}=-\frac{3}{2} m^2 g_{\mu\nu}~,
\ee
and the unique solution preserving all supersymmetry is
$AdS_7$ which we can write in Poincar\'e co-ordinates as
\be
ds^2=\frac{4}{m^2}\left[
\frac{d\xi^i d\xi^j \eta_{ij}+d\rho^2}{\rho^2}\right]~.
\ee
If we uplift this solution to D=11 using \p{d7redform} we recover the
$AdS_7\times S^4$ solution \p{fivenear}, 
with the radius of the $AdS_7$ given by $2/m$.
Note, in these co-ordinates, the D=7 
metric clearly displays the flat planar world-volume of the fivebrane.

\subsection{Fivebranes wrapping SLAG 3-cycles}
Supergravity solutions describing fivebranes wrapping different
calibrated cycles have been constructed in D=7 gauged supergravity 
and then uplifted to D=11 in 
\cite{Maldacena:2000mw,Acharya:2000mu,Gauntlett:2000ng,Gauntlett:2001jj}. 
We will illustrate in some
detail the construction of fivebranes wrapping SLAG 3-cycles \cite{Gauntlett:2000ng}
and then comment more briefly on the other cases.

Consider the D=11 supersymmetric geometry 
$\bbR^{1,2}\times CY_3\times \bbR^2$ and $G=0$,
with a probe fivebrane wrapping a SLAG 3-cycle 
inside the Calabi-Yau three-fold. 
i.e. the world-volume of the fivebrane is $\bbR^{1,2}\times \Sigma_3$ with
$\Sigma_3\subset CY_3$. The five directions transverse 
to the fivebrane world-volume consist of three that are tangent to the $CY_3$
and two flat directions that are normal to the $CY_3$. 
If we wrap many fivebranes, the back reaction on
the geometry will be significant and we aim to find the 
corresponding supergravity solutions with $G\ne 0$.

To construct these solutions, we seek a good ansatz for D=7 gauged supergravity.
By analogy with the flat planar fivebrane solution, 
we think of the D=7 co-ordinates as being those of the world-volume of the fivebrane
plus an additional radial direction which, when the solution
is uplifted to D=11, should correspond to a kind of radial 
distance away from the
fivebrane in the five transverse directions. Thus an obvious ansatz for the D=7 metric is
\be
ds^2=e^{2f}[ds^2(\bbR^{1,2})+dr^2]+e^{2g}d\bar s^2(\Sigma_3)~,
\ee
where $f,g$ are functions of $r$ only and $d\bar s^2(\Sigma_3)$ is some metric
on the 3-cycle. Now before the back reaction
is taken into account, three of the directions transverse to the fivebrane 
were tangent to $CY_3$ and two were flat. We thus decompose the
$SO(5)$-symmetry of the D=7 theory into $SO(3)\times SO(2)$ and only switch on 
$SO(3)$ gauge-fields. 
In order to preserve supersymmetry, as we shall elaborate on shortly, the
$SO(3)$ gauge-fields are chosen to be proportional to the
spin connection of $\Sigma_3$:
\be\label{keyansatz}
2mB^a{}_b=\bar \omega^{a}{}_b~,
\ee
for $a,b=1,2,3$. This is a key part of the ansatz and
it precisely corresponds to the fact that the normal
bundle of SLAG-cycles is isomorphic to the tangent bundle of the
cycle, as we discussed in section 4.1 and also at the end of section 4.3.
An ansatz for the scalar fields respecting $SO(3)\times SO(2)$ 
symmetry is given by
\be
\Pi_A{}^i=diag(e^{2\lambda},e^{2\lambda},e^{2\lambda},e^{-3\l},e^{-3\l})~,
\ee
where $\l$ is a third function of $r$. 
It is consistent with the equations of motion
to set the three-forms $S$ to zero, which we do to complete the ansatz.

{}For the configuration to preserve supersymmetry, we 
require that there exist spinors $\e$ such that
the supersymmetry variations \p{susytran} vanish. 
Given the above ansatz, the composite gauge fields 
$Q$ appearing in \p{nabla} are
given by the $SO(3)$ gauge fields $Q^{ab}=2mB^{ab}$. 
As a consequence, in demanding the vanishing of the variation of 
the gravitini in
the directions along $\Sigma_3$, we find that
\be\label{conds}
(\partial+\frac{1}{4}{\bar\omega}^{ab}\gamma_{ab}+\frac{m}{2} {B}^{ab}\Gamma_{ab})\epsilon=0~,
\ee 
where $\bar\omega$ is the $SO(3)$ spin connection on $\Sigma_3$. We now begin to see
significance of the assumption
\p{keyansatz}. In particular, in the obvious orthonormal frame,
\p{conds} can be satisfied by spinors independent of the 
coordinates along $\Sigma_3$, if we impose the following projections on $\e$:
\be\label{consl3}
\gamma^{ab}\e=-\Gamma^{ab}\e~.
\ee
Note that this precisely parallels our discussion on the preservation of
supersymmetry in the context of the world-volume of the probe fivebrane 
at the end of section 4.3.
To ensure that the variation 
of all components of the gravitini and dilatini vanish we also need to impose
\be\label{rproj}
\gamma^r\e=\e
\ee
and we find that the the only dependence of the Killing spinors on the co-ordinates is radial:
$\e=e^{f/2}\e_0$ where $\e_0$ is a constant spinor. 
Note that only two of the conditions \p{consl3} 
are independent and they break 1/4 of the supersymmetry. When combined with
\p{rproj} we see that 1/8 of the supersymmetry is preserved in agreement with
table 4. Actually, if one follows through these preserved supersymmetries to D=11
one finds that they are precisely those that one expects when a fivebrane wraps
a SLAG 3-cycle: there are two projections corresponding to the $CY_3$ and another
for the fivebrane. 

A more detailed analysis of the conditions for supersymmetry 
implies that the metric on $\Sigma_3$ must in fact be Einstein. 
Given the factor $e^{2g}$, we can always normalise so
that Ricci tensor and the metric on $\Sigma_3$ are related
by 
\be\label{econd}
\bar Ric(\Sigma_3)=l\bar g(\Sigma_3)~,
\ee
with $l=0\pm 1$. In three dimensions the Riemann curvature tensor is determined
by the Ricci tensor. When $l=0$ \p{econd} implies that $\Sigma_3$ is flat:
the resulting D=11 solution, after uplifting, 
corresponds to a fivebrane with a flat planar world-volume and
is thus not of primary interest. Indeed the solution turns out to
be just a special case of the fivebrane solution presented in \p{fivegeom}
with a special harmonic function with $SO(3)\times SO(2)$ symmetry.

The cases of most interest are thus when $l=\pm 1$. 
When $l=1$ the Einstein condition \p{econd} implies that
$\Sigma_3$ is the three-sphere, $S^3$,  or a quotient by a discrete
subgroup of isometries of the isometry group $SO(4)$, while for 
$l=-1$ it is hyperbolic three-space, $H^3$, or a quotient by a discrete
subgroup of $SO(3,1)$. Note in particular that when
$l=-1$ it is possible that the resulting geometry is compact.
We mentioned above that the Killing spinors 
are independent of the coordinates on
the cycle, and hence the quotients $S^3/\Gamma$ and $H^3/\Gamma$ 
are also supersymmetric. 

Finally, in addition, supersymmetry implies the following
first order BPS equations on the three radial functions:
\bea
e^{-f}f'&=& - \frac{m}{10}\left[3e^{-4\lambda}+2e^{6\lambda}\right]
+\frac{3l}{20m }{e^{4 \lambda-2g}}\nn
e^{-f}{g'} &=& - \frac{m}{10}\left[3e^{-4\lambda}+2e^{6 \lambda}\right]
- \frac{7l}{20m }{e^{4 \lambda-2g}}\nn
e^{-f}\lambda' &= & \frac{m}{5}\left[
e^{6\lambda} - e^{-4\lambda}\right] +\frac{l}{10m } {e^{4\lambda-2g}}~.
\eea     
If these equations are satisfied then all of the D=7 equations of motion 
are satisfied and we have found a supersymmetric solution.

Using \p{d7redform} we can now uplift any solution of these BPS equations 
to obtain supersymmetric solutions of D=11 supergravity. The metric is given by
\be\label{slagthreeup}
ds^2_{11}=\Delta^{-{2}/{5}}ds^2_{7} +\frac{1}{m^2}\Delta^{{4}/{5}}
\left[e^{4\lambda}DY^aDY^a+e^{-6\lambda}dY^\a dY^\a\right]~,
\ee
where
\bea\label{slagthreeuptwo}
DY^a&=&dY^a+\bar\omega^{a}{}_{b}Y^b\nn
\Delta^{-{6}/{5}}&=&e^{-4\lambda}Y^aY^a+e^{6\lambda}Y^\a Y^\a~,
\eea
with $a,b=1,2,3$, $\a=4,5$ and $(Y^a,Y^\a)$ are constrained coordinates
on $S^4$ satisfying $Y^aY^a+Y^\a Y^\a=1$. The expression for the four-form
is given by substituting into \p{d7redform}. 
Clearly, the four sphere is no longer round and 
it is non-trivially fibred over the three-dimensional Einstein space $\Sigma_3$.
                                     
It is illuminating\footnote{These type of coordinates were first noticed
in the context of wrapped membranes in \cite{Gauntlett:2001qs}.} 
to change coordinates from the constrained co-ordinates
$(r, Y^A)$ to unconstrained co-ordinates $(\rho^a,\rho^\a)$ via
\bea
\rho^a&=&-\frac{1}{m}e^{f+g+2\l} Y^a\nn
\rho^\a&=&-\frac{1}{m}e^{2f-3\l} Y^\a~.
\eea
The metric then takes the form:
\bea
&ds^2=(\Delta^{-2/5}e^{2f})\left[ds^2(\bbR^{1,2})+e^{2g-2f}d\bar s^2(\Sigma_3)\right]\nn
&\qquad\qquad
+(\Delta^{4/5}e^{-4f})\left[e^{2f-2g}(d\rho^a+\bar\omega^a{}_b \rho^b)^2+d\rho^\a d\rho^\a
\right]~.
\eea 
In these coordinates the warp factors have a similar form 
to the simple planar fivebrane solution \p{fivegeom}; in particular this form confirms
the interpretation of the solutions as describing fivebranes with world-volumes 
given by $\bbR^{1,2}\times \Sigma_3$. In addition the metric clearly displays 
the $SO(2)$ symmetry corresponding to rotations in the $\rho^4,\rho^5$ plane. 

Of course to find explicit solutions we need to solve the BPS equations.
When $l=-1$ it is easy to check that there is an exact solution given by
\bea\label{stscft}
e^{10\l}&=&2\nn
e^{2g}&=&\frac{e^{8\l}}{2m^2}\nn
e^{2f}&=&\frac{e^{8\l}}{m^2 r^2}~.
\eea
The D=7 metric is then the direct product $AdS_4\times (H^3/\Gamma)$ (this D=7 solution
was first found in \cite{Pernici:1985nw}). The uplifted 
D=11 solution is a warped product of $AdS_4\times (H^3/\Gamma)$ with a 
four-sphere which is non-trivially fibred over $(H^3/\Gamma)$. 
The presence of the $AdS_4$ factor in the D=11 solution indicates that M-theory 
on this background is dual to a superconformal field theory
in three spacetime dimensions. We will return to this point after
analysing the general BPS solution. Note that when $l=1$ there is
no $AdS_4\times (S^3/\Gamma)$ solution.

It seems plausible that the BPS equations could be solved exactly.
However, much of the physical content can be deduced from a 
simple numerical investigation. If we introduce the new variables
\bea
\newa^2&=&e^{2g}e^{-8\lambda}\nn
e^\newf&=&e^{f-4\lambda}~,
\eea
then the BPS equations are given by
\bea
e^{-\newf}\newf'&=& - \frac{m}{2}\left[2e^{10\lambda}-1\right]
-\frac{l}{4m\newa^2} \nn
e^{-\newf}\frac{\newa'}{\newa} &=& - \frac{m}{2}\left[2e^{10\lambda}-1\right]
- \frac{3l}{4m\newa^2}\nn
e^{-\newf}\lambda' &= & \frac{m}{5}\left[ 
e^{10\lambda} - 1\right] +\frac{l}{8m\newa^2}~.
\eea
We next define $x=\newa^2$ and $F=xe^{10\lambda}$ to obtain the ODE
\be
\frac{dF}{dx}=\frac{F[m^2x-5\alpha+2\beta]}{x[m^2(2F-x)+2\beta]}~.
\ee

\begin{figure}[!htb]
\vspace{5mm}
\begin{center}
\epsfig{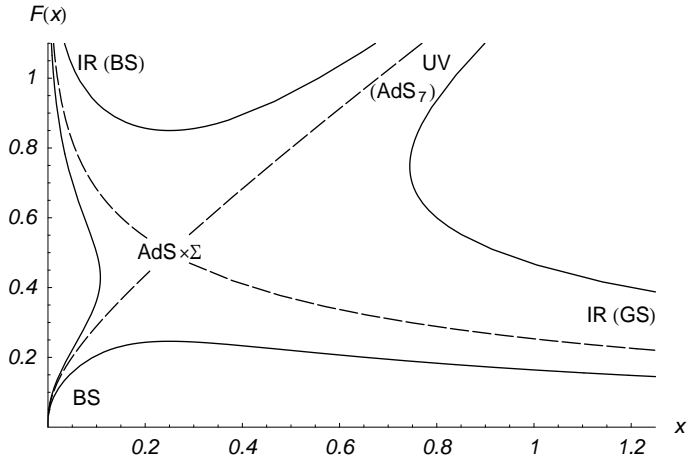}\\
\end{center}
\caption{Behaviour of the orbits for five-branes wrapping SLAG 3-cycles
with $l=-1$. Note the flow from the $AdS_7$-type region when $F,x$ are large to
the IR fixed point and the flows to the good and bad singularities in the IR,
IR(GS) and IR(BS), respectively.}
\label{fig3}
\vspace{5mm}
\end{figure}   
\begin{figure}[!htb]
\vspace{5mm}
\begin{center}
\epsfig{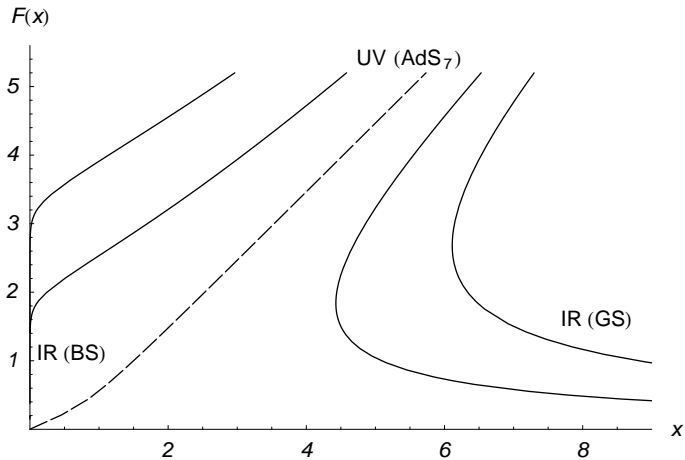}\\
\end{center}
\caption{Behaviour of the orbits for fivebranes wrapping SLAG 3-cycles with $l=1$.}
\label{fig4}
\vspace{5mm}
\end{figure}  

The typical behaviour of $F(x)$ is illustrated in figure 1 for $l=-1$
and figure 2 for $l=1$. The region where both
$x$ and $F$ are large is interesting. There we have
$F\approx x-l/m^2$ and using $\newa$ as a radial variable
we obtain the asymptotic behaviour of the metric:
\be\label{asympt}
ds^2\approx\frac{4}{m^2\newa^2}d\newa^2 +\newa^2[ds^2(\bbR^{1,2})+d\bar{s}^2(\Sigma_3)]\nn~.
\ee
This looks very similar to $AdS_7$ in Poincare co-ordinates
except that the sections with constant $a$ are not $\bbR^{1,5}$ but 
$\bbR^{1,2}\times \Sigma_3$. 

This clearly corresponds to the near horizon limit of the fivebrane wrapped on the
SLAG 3-cycle. By the general discussion on the AdS/CFT correspondence earlier,
this should be dual to the six dimensional quantum field theory living on the 
wrapped fivebrane worldvolume $\bbR^{1,2}\times \Sigma_3$, 
after decoupling gravity. More precisely, the asymptotic behaviour of
the solution \p{asympt}, when lifted to D=11, is dual to the UV behaviour of the 
quantum field theory. Following the flow of the solution as in figures 1 and 2
correspond to flowing to the IR of the field theory. 
In the present context, the IR corresponds to length scales large compared 
to the size of the cycle $\Sigma_3$ on which the fivebrane is wrapped. 
In other words, going to the IR corresponds to taking 
$\Sigma_3$ to be very small (assuming it is compact) and the six dimensional quantum 
field theory on $\bbR^{1,2}\times \Sigma_3$ behaves more and more 
like a three-dimensional quantum field theory on $\bbR^{1,2}$.

Perhaps the most interesting solutions occur for $l=-1$. There is a solution
indicated by one of the dashed lines in figure 1 that flows from 
the UV $AdS_7$ type region to the $AdS_4\times (H^3/\Gamma)$ fixed point that was 
given in \p{stscft}. This is a supergravity solution that describes
a kind of renormalisation group flow from a theory on 
$\bbR^{1,2}\times (H^3/\Gamma) $ to a superconformal field theory on $\bbR^{1,2}$. 
In particular, we see that the natural interpretation of the $AdS_4\times (H^3/\Gamma)$
solution is that, when it is lifted to D=11, it is dual to a superconformal
field theory on $\bbR^{1,2}$ that arises as the IR limit of the fivebrane field
theory on living on $\bbR^{1,2}\times (H^3/\Gamma)$. The interpretation for
non-compact $H^3/\Gamma$ is less clear.

The absence of an $AdS_4\times (S^3/\Gamma)$ solution for $l=1$ possibly
indicates that the quantum field theories arising on fivebranes wrapping SLAG 3-cycles 
with positive curvature are not superconformal in the infra-red. Alternatively, 
it could be that there are more elaborate solutions lying outside of
our ansatz that have 
$AdS_4$ factors.

All other flows in figures 1 and 2 starting from the $AdS_7$ type region,
flow to singular solutions. Being singular does not exclude 
the possibility that they might be interesting physically. 
Indeed, a  criteria for time-like singularities in static geometries
to be ``good singularities'', i.e. dual to some quantum field theory behaviour,
was presented in \cite{Maldacena:2000mw}.
In particular, a good singularity is defined to be one in which 
the norm of the time-like Killing vector with respect to the 
D=11 supergravity metric does not increase as one goes to the singularity (one can also
consider the weaker criteria that the norm is just bounded from above).
It is not difficult to determine whether the singularities that arise
in the different asymptotic limits are good or bad by this criteria
and this has been presented in figures
1 and 2. It is likely that the good singularities describe some kind of Higgs 
branches of the quantum field theory corresponding to the possibility of 
moving the co-incident wrapped fivebranes apart.

In summary, using D=7 gauged supergravity, we have been able to construct 
D=11 supergravity solutions that describe fivebranes wrapping SLAG 3-cycles.
The cycle is Einstein and is either $S^3/\Gamma$ or $H^3/\Gamma$ where $\Gamma$ is
a discrete group of isometries. Probably the most important solutions
that have been found are the $AdS_4\times (H^3/\Gamma)$ solutions which
are dual to new superconformal field theories, after being uplifted to D=11.
More general flow solutions were also constructed numerically.

\subsection{Fivebranes wrapping other cycles}

The construction of supersymmetric solutions corresponding
to fivebranes wrapping other supersymmetric cycles runs along
similar lines. The ansatz for the D=7 metric is given by
\be\label{metansatz}
ds^2=e^{2 f}[ds^2(\bbR^{1,5-\dcy}) +d r^2] + e^{2g}d\bar{s}^2(\Sigma_\dcy)~,
\ee
where $d\bar s^2_\dcy$ is the metric on the supersymmetric
$\dcy$-cycle, $\Sigma_d$, and the functions $f$ and $g$ depend on the radial
coordinate $r$ only. 

The $SO(5)$-gauge fields are specified by the spin connection of the
metric on $\Sigma_\dcy$ in a way determined by the structure
of the normal bundle of the calibrated cycle being wrapped.
In general, we decompose the $SO(5)$ symmetry into 
$SO(p)\times SO(q)$ with $p+q=5$, and only excite the gauge
fields in the $SO(p)$ subgroup. We will denote these directions by
$a,b=1,\dots,p$. If we consider a probe fivebrane wrapping the
cycle inside a manifold of special holonomy $M$,
this decomposition corresponds to dividing the directions
transverse to the brane into $p$ directions within $M$ and $q$
directions perpendicular to $M$. The precise ansatz for the
$SO(p)$ gauge-fields is determined by some part of the spin
connection on the cycle and will be discussed shortly.

In keeping with this decomposition, the solutions that we
consider have a single scalar field excited. More precisely we take
\be\label{scalaransatz}
{\Pi_A}^i=diag(e^{q\lambda},\dots,e^{q\lambda},
e^{-p\lambda},\dots,e^{-p\lambda})~,
\ee
where we have $p$ followed by $q$ entries.
Once again this implies that the composite gauge-field $Q$ is then
determined by the gauge-fields via $Q^{ab}=2m B^{ab}$.

{}It turns out that for the SLAG 5-cycle and most of the 4-cycle cases 
it is necessary to have non-vanishing three-forms $S_A$.
The $S$-equation of motion is 
\be\label{seom}
m^2\delta_{AC}{{\Pi^{-1}}_i}^C{{\Pi^{-1}}_i}^BS_{B}
=-m*F_A+\frac{1}{4\sqrt 3}\epsilon_{ABCDE}*(F^{BC}\wedge F^{DE})~.
\ee
The solutions have vanishing four-form field strength $F_A$ and
hence $S_A$ are determined by the gauge fields.

Demanding that the configuration preserves supersymmetry,
as for the SLAG 3-cycle case, we find that along the 
cycle directions:
\be\label{condsagain}
(\partial+\frac{1}{4}{\bar\omega}^{bc}\gamma_{bc}+\frac{m}{2} {B}^{ab}\Gamma_{ab})\epsilon=0~,
\ee
where ${\bar\omega}^{bc}$ is the spin connection one-form of the
cycle. For each case the specific gauge-fields, determined by the
type of cycle being wrapped, go hand in hand with a set of projections
which allow \p{condsagain} to be satisfied for spinors independent of the coordinates
along $\Sigma_d$. We can easily guess the appropriate 
ansatz for the gauge fields from
our discussion of the structure of the normal bundles of 
calibrated cycles in section 4.1, and they are summarised below.
The corresponding projections are then easily determined and are given 
explicitly in 
\cite{Maldacena:2000mw,Acharya:2000mu,Gauntlett:2000ng,Gauntlett:2001jj}.
In the uplifted D=11 solutions, these projections translate into a 
set of projections corresponding to those of the special holonomy manifold 
and an additional projection corresponding to the wrapped fivebrane.

By analysing the conditions for supersymmetry in more detail, one finds
that the metric on the cycle is necessarily Einstein, and we again
normalise so that
\be\label{ein}
\bar R_{ab}=l \bar g_{ab}~,
\ee
with $l=0,\pm 1$. When $d=2,3$ the 
Einstein condition implies that the cycles have constant curvature and hence
are either spheres, for $l=1$ or hyperbolic spaces for $l=1$, or quotients
of these spaces by a discrete group of isometries.
When $d>3$ the Einstein condition implies that the Riemann 
tensor can be written
\be\label{decom}
\bar R_{abcd}=\bar W_{abcd} +\frac{2l}{\dcy -1}\bar g_{a[c}\bar g_{d]b}~,
\ee
where ${\bar W}$ is the Weyl tensor, and there are more possibilities.
By analysing the D=7 Einstein equations one finds that for $d=4,5$ 
the part of the spin connection that is identified with
the gauge fields must have constant curvature.

We now summarise the ansatz for the $SO(5)$ gauge fields for each case
and discuss the types of cycle that arise for $=4,5$. 
We can always take a quotient of the cycles listed by a discrete 
group of isometries.

{\bf SLAG $n$-cycles}: Consider a probe fivebrane wrapping a SLAG $n$-cycle in
a $CY_n$. The five directions transverse to the fivebrane consist of $n$ directions
tangent to the $CY_n$ and $5-n$ normal to it. Thus, for the supergravity solution
we decompose $SO(5)\to SO(n)\times SO(5-n)$, and let the only non-vanishing gauge
fields lie in the $SO(n)$ factor. Up to a factor of $2m$ these $SO(n)$ gauge-fields
are identified with $SO(n)$ spin connection on $\Sigma_n$, as in \p{keyansatz}. This identification
corresponds to the fact that 
$N(\Sigma_n)\cong T(\Sigma_n)$ for SLAG $n$-cycles. Since all of the
spin connection is identified with the gauge-fields,  
the metric on $\Sigma_n$ must have constant curvature ($W=0$ for $d=4,5$) and 
hence $\Sigma_n$ is $S^n$ for $l=1$ and $H^n$ for $l=-1$. 

{\bf K\"ahler 2-cycles}: K\"ahler 2-cycles in $CY_2$ are SLAG 2-cycles and have just
been discussed. For probe fivebranes wrapping K\"ahler 2-cycles in $CY_3$ 
the five directions transverse to the fivebrane 
consist of four directions tangent to the 
$CY_3$ and one flat direction normal to the $CY_3$.
The normal bundle of the K\"ahler 2-cycle has structure group 
$U(2)\cong U(1)\times SU(2)$ and recall now \p{coneid}.
Thus, for the supergravity solution in D=7, 
we decompose $SO(5)\to SO(4)\to U(2)\cong U(1)\times SU(2)$ 
and identify the $U(1)$ spin connection of the cycle with the gauge-fields
in the $U(1)$ factor. 
We also set the $SU(2)$ gauge-fields to zero, which corresponds to considering fivebranes
wrapping K\"ahler 2-cycles with non-generic normal bundle. An example of such a
cycle is the two-sphere in the resolved conifold.
It would be interesting to find more general solutions with non-vanishing $SU(2)$ gauge-fields,
preserving the same amount of supersymmetry.

{\bf K\"ahler-4-cycles}: We assume that $\Sigma_4$ in the D=7 supergravity solution
has a K\"ahler metric with a $U(2)\cong U(1)\times SU(2)$ 
spin connection.
When probe fivebranes wrap K\"ahler 4-cycles in $CY_3$ 
the five transverse directions consist of two directions tangent to the $CY_3$ 
and three flat directions normal to the $CY_3$.  Thus we decompose 
$SO(5)\to SO(2)\times SO(3)$, set the $SO(3)$ gauge-fields to zero 
and identify the $SO(2)\cong U(1)$ gauge fields with the 
$U(1)$ part of the $U(1)\times SU(2)$ spin connection on the 4-cycle. 
When probe fivebranes wrap K\"ahler 4-cycles in $CY_4$ 
the five transverse directions consist of four directions tangent to the 
$CY_4$ and one flat direction normal to the $CY_4$. 
Thus we now decompose $SO(5)\to SO(4)\to
U(2)\cong U(1)\times SU(2)$ and we set the $SU(2)$ gauge-fields to
zero, which again corresponds to considering non-generic normal
bundles. We identify the $U(1)$ gauge-fields with
the $U(1)$ part of the spin connection as dictated by \p{coneid}.
In both cases, the identification of the gauge fields with part of the spin
connection doesn't place any further
constraints on $\Sigma_4$ other than it is K\"ahler-Einstein. 
An example when $l=1$ is $CP^2$.

{\bf $\bbC$-Lag 4-cycles}: 
We again assume that $\Sigma_4$ in the D=7 solution
has a K\"ahler metric with a $U(2)\cong U(1)\times SU(2)$ 
spin connection. We again decompose $SO(5)\to SO(4)\to
U(2)\cong U(1)\times SU(2)$ but now we do not
set the $SU(2)$ gauge-fields to zero. Indeed, since the
cycle is both SLAG and K\"ahler, with respect to different complex
structures, we must identify all of the
$U(2)$ gauge-fields with the $U(2)$ spin connection.
Einstein's equations then imply that $\Sigma_4$ must
have constant holomorphic sectional curvature. This
means that for $l=1$ it is $CP^2$ while for $l=-1$ it is the open disc in $\bbC^2$
with the Bergman metric. Note that the solutions corresponding to fivebranes
wrapping $\bbC$-Lag $CP^2$ are different from the solutions corresponding to 
fivebranes wrapping K\"ahler $CP^2$, since they have more gauge-fields excited and
preserve different amounts of supersymmetry.

{\bf Associative 3-cycles}: When probe fivebranes wrap associative 3-cycles in
$G_2$ manifolds, there are four transverse directions that are tangent to the $G_2$
manifold and one flat direction normal to the $G_2$ manifold. We thus decompose
$SO(5)\to SO(4)\cong SU(2)^+\times SU(2)^-$, where the superscripts indicate
the self-dual and anti-self-dual parts. Recall that the normal bundle
of associative 3-cycles is given by $S\otimes V$ where $S$ was the $SU(2)$
spin bundle on $\Sigma_3$ and $V$ is a rank $SU(2)$ bundle. In the
non-generic case when $V$ is trivial, for example for the $G_2$ manifold
in \cite{MR90i:53055}, then the identification of the gauge-fields is clear: we should
identify the $SO(3)\cong SU(2)$ spin connection on $\Sigma_3$ with
$SU(2)^+$ gauge-fields and set the $SU(2)^-$ gauge-fields to zero. 

{\bf Co-associative 4-cycles}:  When probe fivebranes wrap co-associative 4-cycles in
$G_2$ manifolds, there are three transverse directions that are tangent to the $G_2$
manifold and two flat directions normal to the $G_2$ manifold. We thus decompose
$SO(5)\to SO(3)\times SO(2)$ and set the $SO(2)$ gauge-fields to zero.
Recall that the normal bundle of co-associative 4-cycles 
is isomorphic to the bundle of anti-self-dual two-forms on the 4-cycle.
This indicates that we should identify the $SO(3)\cong SU(2)$ gauge-fields
with the anti-self-dual part, $SU(2)^-$, of the 
$SO(4)\cong SU(2)^+\times SU(2)^-$ spin connection on $\Sigma_4$.
{}For the co-associative 4-cycles, Einstein's equations imply that
the anti-self-dual part of the spin connection has constant curvature, or
in other words, the Weyl tensor is self-dual $W^-=0$. These manifolds
are sometimes called conformally half-flat.
If $l=1$ the only compact examples are  $CP^2$ and $S^4$.

{\bf Cayley 4-cycles}: When probe fivebranes wrap Cayley 4-cycles in $Spin(7)$ 
manifolds the five normal directions consist of four directions tangent to the 
$Spin(7)$ manifold and one flat direction normal to the $Spin(7)$ manifold. 
We thus again decompose
$SO(5)\to SO(4)\cong SU(2)^+\times SU(2)^-$, where the superscripts indicate
the self-dual and anti-self-dual parts.
Recall that the normal bundle of Cayley 4-cycles 
is given by $S_-\otimes V$ where $S_-$ was the $SU(2)$
bundle of negative chirality spinors on $\Sigma_3$ 
and $V$ is a rank $SU(2)$ bundle. In the
non-generic case when $V$ is trivial, for example for the $Spin(7)$ manifolds
in \cite{MR90i:53055}, then the identification of the gauge-fields is clear: 
if $SU(2)^\pm$ are the self-dual and anti-self-dual parts of the
$SO(4)$ spin connection on $\Sigma_4$ then we should identify
the $SU(2)^-$ part of the spin connection with $SU(2)^-$ gauge fields
and set the $SU(2)^+$ gauge-fields to zero.
As for the co-associative 4-cycles, Einstein's equations imply that
the Weyl tensor is self-dual $W^-=0$.

With this data the BPS equations can easily be derived and we refer to 
\cite{Maldacena:2000mw,Acharya:2000mu,Gauntlett:2000ng,Gauntlett:2001jj}
for the explicit equations. 
They have a similar appearance to those of the SLAG 3-cycle case,
with the addition of an extra term coming from the three-forms $S$ for
most of the 4-cycle cases and the SLAG 5-cycles. 
When the curvature of the cycle is negative, $l=-1$, in all cases except 
for K\"ahler 4-cycles in $CY_3$, we obtain an $AdS_{7-d}\times \Sigma_d$ 
fixed point. When $l=1$, only for SLAG 5-cycles do we find 
such a fixed point. This
is summarised in table 5.

We note that using exactly the same ansatz for the D=7 supergravity fields,
some additional non-supersymmetric solutions of the form 
$AdS_{7-d}\times \Sigma_{d}$ were found \cite{Gauntlett:2002rv}.
In addition, by considering the possibility of extra scalar fields being excited,
one more $AdS_3$ solution was found.
Such solutions could be dual to non-supersymmetric conformal field theories, that are
related to wrapped fivebranes with supersymmetry broken. To develop
this interpretation it is necessary that the solutions are stable, which is
difficult to determine. A preliminary perturbative investigation revealed that 
some of these solutions are unstable. We have also summarised these solutions
in table 5.

\begin{table}[htbp]
\begin{center}
\begin{tabular}{|c|c|c|c|}
   \hline
   spacetime & embedding & cycle $\Sigma_n$ & supersymmetry  \\ 
   \hline
   \hline
   $AdS_5\times \Sigma_2$ & K\"{a}hler 2-cycle in $CY_2$ & $H^2$ & yes \\ 
           &                              & $S^2$ & no$^*$   \\
   \cline{2-4}
           & K\"{a}hler 2-cycle in $CY_3$ & $H^2$ & yes  \\
           &                              & $H^2$ & no  \\
   \hline
   $AdS_4\times \Sigma_3$ & SLAG 3-cycle in $CY_3$ & $H^3$ & yes  \\ 
           &                        & $H^3$ & no    \\
   \cline{2-4}
           & Associative 3-cycle    & $H^3$ & yes  \\
           &                        & $H^3$ & no   \\
   \hline
   $AdS_3\times \Sigma_4$ & Coassociative 4-cycle  & $C_-^4$ & yes  \\ 
   \cline{2-4}
           & SLAG 4-cycle in $CY_4$ & $H^4$   & yes  \\
           &                        & $H^4$   & no   \\
           &                        & $S^4$   & no   \\
   \cline{2-4}
           & K\"{a}hler 4-cycle in $CY_4$ & $K_-^4$ & yes  \\
           &                              & $K_+^4$ & no   \\
   \cline{2-4}
           & Cayley 4-cycle & $C_-^4$ & yes  \\
           &                & $C_-^4$ & no   \\
           &                & $CP^2, S^4$ & no   \\
   \cline{2-4}
           & CLAG 4-cycle in $HK_8$ & $B$    & yes  \\
           &                        & $B$    & no   \\
           &                        & $CP^2$ & no   \\
   \cline{2-4}
           & SLAG 4-cycle in $CY_2\times CY_2$ & $H^2\times H^2$ 
                                     & yes \\
           &       & $H^2\times H^2$ & no$^*$   \\
           &       & $S^2\times S^2$ & no$^*$   \\
           &       & $S^2\times H^2$ & no$^*$   \\
   \hline
   $AdS_2\times \Sigma_5$ & SLAG 5-cycle in $CY_5$ & $H^5$ & yes  \\
           &                        & $S^5$ & yes  \\
   \cline{2-4}
           & SLAG 5-cycle in $CY_2\times CY_3$ & $H^3\times H^2$ 
                                     & yes  \\
           &       & $S^2\times H^3$ & no   \\
           &       & $S^2\times H^3$ & no   \\
   \hline
\end{tabular}
   \caption{$AdS$ fixed point solutions for wrapped fivebranes: 
   $C_-$ and $K_\pm$ are conformally half-flat and
   K\"{a}hler--Einstein metrics with the subscript denoting
   positive or negative scalar
   curvature and $B$ is the Bergmann metric. Note that we can also
   take quotients of all cycles by discrete groups of isometries
   and this preserves supersymmetry. $^*$ denotes a solution shown to be unstable.} 
   \label{tab:tablesum}
\end{center}
\end{table}

\subsection{Wrapped Membranes and D3-branes}

D=11 supergravity solutions describing membranes wrapping K\"ahler 2-cycles
can be found in an analogous manner \cite{Gauntlett:2001qs}. 
The appropriate gauged supergravity for this case is maximal $SO(8)$ gauged supergravity 
in D=4 \cite{deWit:1982ig,deWit:1982eq}
which can be obtained from a consistent truncation of the
dimensional reduction of D=11 supergravity on a seven sphere 
\cite{deWit:1985nz,deWit:1987iy}. 
The vacuum solution of this theory is $AdS_4$ and this uplifts 
to $AdS_4\times S^7$, which is the near horizon limit of the planar
membrane solution. More general solutions can be found that
uplift to solutions describing the near horizon limit of wrapped membranes.

Actually, the general formulae for obtaining $SO(8)$ gauged supergravity from
the dimensional reduction of D=11 supergravity on the seven sphere are 
rather implicit and not in a form that is useful for uplifting general solutions. 
Luckily, there is a further consistent truncation of the 
$SO(8)$ gauged supergravity theory to
a $U(1)^4$ gauged supergravity where the formulae are known explicitly 
(in the special case that the axion fields are zero) \cite{Cvetic:1999xp} 
and this
is sufficient for the construction of D=11 wrapped membrane solutions.

To see why, let us describe the ansatz for the gauge-fields for the D=4 solutions.
If we consider a probe membrane wrapping a K\"ahler 2-cycle in a $CY_n$ then the eight
directions transverse to the membrane consist of $2n-2$ directions that are
tangent to the $CY_n$ and $10-2n$ flat directions that are normal to the $CY_n$.
In addition the normal bundle of the K\"ahler 2-cycle in $CY_n$ has structure group 
$U(n-1)\cong U(1)\times SU(n-1)$ and recall \p{coneid}.
Thus, in the D=4 supergravity, we should first decompose $SO(8)\to SO(2n-2)\times SO(10-2n)$
and only have non-vanishing gauge fields in $U(n-1)\subset  SO(2n-2)$. The
gauge-fields in the $U(1)$ factor of $U(n-1)\cong U(1)\times SU(n-1)$ are then identified
with the $U(1)$ spin connection on the 2-cycle, corresponding to \p{coneid}. 
{}In the solutions that have been constructed,
the remaining $SU(n-1)$ gauge-fields are set to zero, which corresponds to the
normal bundle of the K\"ahler 2-cycle being non-generic when $n\ge 3$. Thus, the ansatz
for the gauge fields are such that they always lie within the maximal cartan
subalgebra $U(1)^4$ of $SO(8)$ and hence the truncation formulae of \cite{Cvetic:1999xp}
can be used.

Once again the metric on the 2-cycle is Einstein and hence is either $S^2/\Gamma$ for $l=1$
or $H^2/\Gamma$ for $l=-1$. In particular, the cycle can be an arbitrary Riemann surface.
General BPS equations have been found and analysed numerically. 
Interestingly, $AdS_2\times \Sigma_2$ fixed points are found only for $l=-1$ and only for the
cases of $CY_4$ and $CY_5$. When uplifted to 
D=11 these solutions become a warped product
with a non-round seven sphere that is non-trivially fibred over the cycle.
The $AdS_2$ fixed point solutions should be dual to superconformal quantum mechanics 
living on the wrapped membranes.

The appropriate gauged supergravity theory for finding D=10 type IIB solutions 
describing D3-branes wrapping various calibrated cycles, is the maximally 
supersymmetric $SO(6)$ gauged supergravity in D=5 \cite{Gunaydin:1986cu}. 
This can be obtained from the consistent truncation of the dimensional 
reduction of the type IIB supergravity on a five-sphere. In particular, the vacuum 
solution is $AdS_5$ and this
uplifts to $AdS_5\times S^5$ which is the near horizon limit of the planar D3-brane.
Actually the general formulae for this reduction are not yet known and one
has to exploit further consistent truncations that are known 
\cite{Romans:1986ps,Cvetic:1999xp,Lu:1999bw,Cvetic:2000nc}. 

All cases in table 3 have been investigated, and BPS equations have been 
found and analysed. 
Once again, in the solutions, the 2- and 3-cycles that the 
D3-branes wrap have Einstein metrics and hence have constant curvature. 
D3-branes wrapping K\"ahler 2-cycles in $CY_2$ and $CY_3$ were studied in
\cite{Maldacena:2000mw} while the $CY_4$ case was analysed in \cite{Naka:2002jz}.
$AdS_3\times H^2/\Gamma$ fixed points were found for the $CY_3$ and $CY_4$ cases.
D3-branes wrapping associative 3-cycles were analysed in \cite{Nieder:2000kc} and an 
$AdS_2\times H^3/\Gamma$ fixed point was found. Finally,
D3-branes wrapping SLAG 3-cycles were studied in \cite{Naka:2002jz} and no
$AdS_2$ fixed point was found. 
Note that non-supersymmetric $AdS$ solutions were sought in 
\cite{Naka:2002jz} for both wrapped membranes and D3-branes, generalising 
the fivebrane solutions in \cite{Gauntlett:2002rv}, but none were
found.

\subsection{Other wrapped brane solutions}

Let us briefly mention some other supergravity solutions describing wrapped branes that
have been constructed.

D6-branes of type IIA string theory carry charge under the $U(1)$ gauge-field
arising from the Kaluza-Klein reduction of D=11 supergravity on an $S^1$ (the field
$C^{(1)}$ in \p{iiared}).
The planar D6-brane uplifts to pure geometry: $\bbR^{1,6}\times M_4$ where $M_4$ is 
Taub-NUT space with $SU(2)$ holonomy. Similarly, when D6-branes wrap calibrated cycles
they uplift to other special holonomy manifolds in D=11 and this has been studied
in e.g.
\cite{Acharya:2000gb,Atiyah:2000zz,Gomis:2001vk,Edelstein:2001pu,Hernandez:2001bh,Gomis:2001vg}.
Other solutions related to wrapped D6-branes that are dual to non-commutative field theories
have been studied in \cite{Brugues:2002pm,Brugues:2002ff}.

There are solutions of massive type IIA supergravity with $AdS_6$ factors which
are dual to the five-dimensional conformal field theory arising on the
D4-D8-brane system \cite{Ferrara:1998gv,Brandhuber:1999np}. 
Supersymmetric and non-supersymmetric solutions describing the 
D4-D8 system wrapped on various calibrated cycles were found in
\cite{Nunez:2001pt,Naka:2002jz}.

{}For some other supergravity solutions with possible
applications to AdS/CFT, that are somewhat related to those described here, see 
\cite{Alishahiha:1999ds,Cvetic:2000cj,Fayyazuddin:1999zu,
Fayyazuddin:2000em,Brinne:2000fh,Chamseddine:1999xk,Klemm:2000nj,Chamseddine:2000bk,Klemm:2000vn}.

\section{Discussion}

We have explained in some detail the construction of supergravity solutions
describing branes wrapping calibrated cycles. There are a number of issues 
that are worth further investigation. 

It seems plausible that the BPS equations can be solved exactly. To date
this has only been achieved in a few cases. They were solved for the
case of  membranes wrapping K\"ahler 2-cycles in $CY_5$ \cite{Gauntlett:2001qs}, 
but this case is special in that all scalar fields in the gauged supergravity 
are set to zero. When there is a non-vanishing scalar field, the 
BPS equations were solved exactly for some cases in \cite{Maldacena:2000mw} 
and they have been partially integrated for other cases. 
Of particular interest are the exact solutions 
corresponding to the flows from an $AdS_D$ region to an 
$AdS_{D-d}\times \Sigma_d$ fixed point (for example, one of the dashed 
lines in figure 1) as they are completely regular solutions. 

{}For the case of  membranes wrapping K\"ahler 2-cycles in $CY_5$, 
the general flow solution can be viewed as the ``topological'' $AdS_4$ black holes 
discussed in \cite{Caldarelli:1998hg}. When $l=-1$,
there is a supersymmetric rotating generalisation of this black  
hole \cite{Caldarelli:1998hg}: when it uplifted to D=11, it 
corresponds to waves on the wrapped membrane \cite{Gauntlett:2001qs}. The 
rotating solution
is completely regular provided that the angular momentum is bounded.
It would be interesting to understand this bound from the point of view of the
dual field theory.
In addition, the existence of this rotating solution suggests that for
all of the regular flow solutions of wrapped branes 
starting from an 
$AdS_D$ region and flowing to an $AdS_{D-d}\times \Sigma_d$ region, 
there should be
rotating generalisations that are waiting to be found.

In all of the supergravity solutions describing wrapped branes that
have been constructed, the cycle has an
Einstein metric on it. It would be interesting if a more general ansatz could
be found in which this condition is relaxed. While this seems possible,
it may not be possible to find explicit solutions.
In some cases, such as fivebranes wrapping K\"ahler 2-cycles in $CY_3$, we noted
that the solutions constructed correspond to fivebranes wrapping cycles with
non-generic normal bundles. This was because certain gauge-fields
were set to zero. We expect that more general solutions can be found corresponding to 
generic normal bundles. Note that such a solution, with an AdS factor, 
was found for D3-branes wrapping K\"ahler 2-cycles in $CY_3$
\cite{Maldacena:2000mw}. 

It would also be interesting to construct more general solutions that describe the 
wrapped branes beyond the near horizon limit. Such solutions 
would asymptote to a special holonomy manifold, which would necessarily be non-compact
in order that the solution can carry non-zero flux (a no-go theorem for D=11 
supergravity
solutions with flux is presented in \cite{Gauntlett:2002fz}). 
It seems likely that solutions can be
found that asymptote to the known co-homogeneity-one special holonomy manifolds.
{}For example, the deformed conifold is a co-homogeneity-one $CY_3$ that
is a regular deformation of the conifold. It has a SLAG three-sphere and topologically
the manifold is the co-tangent bundle of the three-sphere, $T^*(S^3)$. 
It should be possible to generalise the solutions 
describing fivebranes wrapping SLAG
three-spheres in the near horizon limit, to solutions that 
include an asymptotic region far from the branes that approaches 
the conifold metric. Of course, these solutions will still be singular in the near
horizon limit. It will be particularly interesting to construct 
similar solutions for the
$l=-1$ case. $T^*(H^3)$ admits a $CY_3$ metric, with a SLAG $H^3$ but there is a singularity at 
some finite distance from the SLAG 3-cycle. There may be a solution with non-zero
flux that interpolates
from this singular behaviour down to the regular near horizon solutions that we
constructed. Alternatively, it may be that the flux somehow ``pushes off'' the singularity
to infinity and the entire solution is regular.
The construction of these more general solutions, when the four-sphere transverse
to the fivebrane is allowed to get large, will necessarily require new
techniques, as they cannot be found in the gauged supergravity.

The construction of the wrapped brane solutions 
using gauged supergravity is rather indirect and
it is desirable to characterise the D=11 geometries more directly. 
{}For example, this may lead to new methods to generalise the solutions along
the lines mentioned above. One approach, is to guess general 
ansatz\"e for D=11 supergravity configurations that might describe wrapped
brane solutions and then impose the conditions to have supersymmetry. 
This approach has its origins in the construction of the 
intersecting brane solutions, reviewed in 
\cite{Gauntlett:1997cv,Smith:2002wn}, and was further extended in e.g. 
\cite{Fayyazuddin:1999zu,Cho:2000hg,Husain:2002tk,Husain:2003df}. 
Recently, it has been appreciated that it is possible to systematically 
characterise supersymmetric solutions of supergravity theories 
with non-zero fluxes using the notion of $G$-structures 
\cite{Gauntlett:2001ur,Friedrich:2001yp,Ivanov:2001ma,
Gauntlett:2002sc,Gauntlett:2002fz,Gauntlett:2003cy} 
(see also \cite{Gauntlett:2002nw,Gauntlett:2003fk}). 
In particular, it was emphasised in some of these works that
generalised calibrations play a central role 
and this is intimately 
connected with the fact that supergravity solutions with non-vanishing
fluxes arise when branes wrap calibrated cycles. 
It should be noted that while these techniques provide
powerful ways of characterising the D=11 geometries it is often difficult
to obtain explicit examples: indeed even recovering the known 
explicit solutions found via gauged supergravity can be non-trivial 
(see e.g. \cite{Gauntlett:2002sc}).

The supergravity solutions can be used to learn a lot about the dual
conformal field theory, assuming the AdS/CFT correspondence is valid. 
For example, the $AdS$ fixed points can be used to determine
the spectrum and correlation functions of the operators in the dual field theory.
For the case of wrapped D3-branes, since the dual
field theory is related to ${\mathcal N}=4$ super Yang-Mills theory,
some detailed comparisons can be made \cite{Maldacena:2000mw}. 
For the fivebrane case it will be more difficult to do this since 
the conformal field theory living on
the fivebrane is still poorly understood. Perhaps some detailed comparisons
can be made for the wrapped membranes.

Recently some new supersymmetric solutions with AdS factors were constructed in 
\cite{Cucu:2003bm,Cucu:2003yk}. It will be interesting to 
determine their dual CFT interpretation and to
see if they are related to wrapped branes.
The non-supersymmetric solutions containing AdS factors found in
\cite{Gauntlett:2002rv} might be dual to non-supersymmetric conformal 
field theories. A necessary requirement is that
the solutions are stable: it would be useful to complete
the preliminary analysis of the perturbative
stability undertaken in \cite{Gauntlett:2002rv}.

Supergravity solutions that are dual to supersymmetric quantum 
field theories that are not conformally invariant can be constructed
using wrapped $NS$ 5-branes of type IIB string theory.
The near horizon limit of the planar $NS$ 5-brane
is dual to what is known as ``little string theory'' in six dimensions (for a review
see \cite{Aharony:1999ks}). 
These still mysterious theories are not local quantum field theories 
but at low-energies they give rise to supersymmetric Yang-Mills (SYM) theory 
in six dimensions. 
As a consequence, the geometries describing $NS$ 5-branes wrapped
on various calibrated cycles encode information about various
SYM theories in lower-dimensions.
The geometries describing $NS$ 5-branes wrapped 
on K\"ahler 2-cycles in $CY_3$ were constructed in
\cite{Chamseddine:2001hk,Maldacena:2000yy} and are dual to ${\mathcal N}=1$ 
SYM theory in four dimensions.
If the 5-branes are wrapped on K\"ahler 2-cycles in $CY_2$ the geometries
are dual to
${\mathcal N}=2$ SYM theory 
in four dimensions \cite{Gauntlett:2001ps,Bigazzi:2001aj} 
(see also \cite{Hori:2002cd}).
By wrapping on associative 3-cycles one finds geometries 
that encode information about
${\mathcal N}=1$ SYM in D=3 
\cite{Acharya:2000mu,Chamseddine:2001hk,Schvellinger:2001ib,Maldacena:2001pb,Gomis:2001xw}, 
while wrapping on SLAG 3-cycles one finds
${\mathcal N}=2$ SYM in D=3 \cite{Gauntlett:2001ur,Gomis:2001aa}.
{}For the latter case, the solutions presented in 
\cite{Gauntlett:2001ur,Gomis:2001aa} are singular and correspond
to vanishing Chern-Simons form in the dual SYM theory. There are strong physical
arguments that suggest there are more general
regular solutions that are dual to SYM with non-vanishing Chern-Simons
form, and it would be very interesting to construct them.
Note that supergravity solutions describing $NS$ 5-branes wrapping various
4-cycles were found in \cite{Naka:2002jz}.
The D=10 geometry for wrapped $NS$ 5-branes has been analysed in some
detail in 
\cite{Friedrich:2001nh,Gauntlett:2001ur,Friedrich:2001yp,Ivanov:2001ma,
Gauntlett:2002sc,Gauntlett:2003cy}.

We hope to have given the impression that while much is now known about
supergravity solutions describing branes wrapped on calibrated cycles,
there is still much to be understood.

\bibliographystyle{amsplain}
%\bibliography{ref} 

\providecommand{\bysame}{\leavevmode\hbox to3em{\hrulefill}\thinspace}
\providecommand{\MR}{\relax\ifhmode\unskip\space\fi MR }
\providecommand{\MRhref}[2]{%
  \href{http://www.ams.org/mathscinet-getitem?mr=#1}{#2}
}
\providecommand{\href}[2]{#2}

\end{document}